\documentclass[preprint]{aastex}
\usepackage{graphicx}

\shorttitle{Structure and Kinematics of Galaxy Thin Gaseous Disk}

\shortauthors{Galazutdinov et al.}

\begin{document}

\title{The structure and kinematics of the the Galaxy thin gaseous disc outside the solar orbit
\thanks{This paper includes data gathered with the 2~m telescope at the
    ICAMER Observatory (Terskol, Russia) and the 1.8 m telescope at Bohyunsan Optical
    Astronomy Observatory (South Korea).
}}

\author{G.~Galazutdinov$^{1,2}$}
\affil{$^1$ Instituto de Astronomia, Universidad Catolica del Norte, Av. Angamos 0610, Antofagasta, Chile}
\affil{$^2$ Pulkovo Observatory, Pulkovskoe Shosse 65, Saint-Petersburg 196140, Russia}
\email{runizag@gmail.com}

\author{A. Strobel}
\affil{Center for Astronomy, Nicholas Copernicus University, Gagarina 11, Pl-87-100 Toru{\'n}, Poland}
\email{astrobel@astri.umk.pl}

\author{F.A.Musaev$^{3,4,5}$}
\affil{$^3$ Special Astrophysical Observatory of the Russian AS, Nizhnij Arkhyz 369167, Russia}
\affil{$^4$ Institute of Astronomy of the Russian AS, 48 Pyatnitskaya st., 119017 Moscow, Russia}
\affil{$^5$ Terskol Branch of Institute of Astronomy of the Russian AS, 361605 Peak Terskol, Kabardino-Balkaria, Russia}
\email{faig@sao.ru}

\author{A. Bondar}
\affil{ICAMER, Peak Terskol, Kabardino-Balkaria, 361605, Russia}
\email{arctur.ab@gmail.com}

\and

\author{J.~Kre{\l}owski }
\affil{Center for Astronomy, Nicholas Copernicus University, Gagarina 11, Pl-87-100 Toru{\'n}, Poland}
\email{jacek@astri.umk.pl}

\begin{abstract}
The rotation curve of the Galaxy is generally thought to be flat.
However, using radial velocities from interstellar molecular
clouds, which is common in rotation curve determination, seems to
be incorrect and may lead to incorrectly inferring that the
rotation curve is flat indeed. Tests basing on photometric and
spectral observations of bright stars may be misleading. The
rotation tracers (OB stars) are affected by motions around local
gravity centers and pulsation effects seen in such early type
objects. To get rid of the latter a lot of observing work must be
involved. We introduce a method of studying the kinematics of the
thin disc of our Galaxy outside the solar orbit in a way that
avoids these problems. We propose a test based on observations of
interstellar Ca{\sc ii} H and K lines that determines both radial
velocities and distances. We implemented the test using stellar
spectra of thin disc stars at galactic longitudes of 135{\degr}
and 180{\degr}. Using this method, we constructed the  rotation
curve of the thin disc of the Galaxy. The test leads to the
obvious conclusion that the rotation curve of the thin gaseous
galactic disk, represented by the Ca{\sc ii} lines, is Keplerian
outside the solar orbit rather than flat.
\end{abstract}


\section{Introduction}

It is very likely that our Galaxy is a spiral barred galaxy, i.e.
its nucleus is surrounded by a bar. The dominant component at the
location of the Sun is the thin disc, composed of young stars and
interstellar clouds, of which the Sun is a tiny part. The interior
of the Galaxy is dominated by the bulge/bar, while further out the
dominant component is believed to be the dark halo, probably
consisting of dark matter (DM), about which we know very little.
The dark halo is especially problematic for a Galaxy modeler
because it can so far only be detected through its gravitational
field. Dark matter is thought to be made up of a one or more types
of elementary particles that have not been detected on the Earth.
The properties of these particles can only be speculated about, so
it is hard to design an experiment to detect them. A major goal of
a Galaxy modeler is to make a map of the gravitational field of
the dark halo from observations. (The mass distribution of the
dark halo can be inferred from the gravitational field that it
generates.) The difference between the gravitational field of the
whole Galaxy and the combined fields of the visible components
yields the field of the dark halo.

Over the last few years a sustained effort has been made to
interpret observations of the kinematics of interstellar gas,
based on the assumption that H{\sc i}, H{\sc ii} and H$_2$ clouds
move on closed orbits, usually assumed to be circular. These
clouds are considered as tracers of the gaseous thin disc. Visual
inspection of UV and visual pictures of galaxies, situated
edge-on, shows that  the cold, likely evolved objects, are
distributed in a different way than the dense interstellar matter
and OB stars that form thin galactic discs. However, the case of
H{\sc i} clouds, being revealed by the 21 cm spectral feature, is
not that certain. It is evident that galaxies were formed out of
neutral hydrogen but other abundant elements have been formed
inside galaxies.

Molecular clouds and ionized hydrogen clouds are more likely to be
spatially correlated with the thin disc since OB stars, which
excite H{\sc ii} clouds and are used to measure their distances,
are recently formed out of dense, molecular, interstellar clouds.
However, molecular spectral features seen in spectra of OB stars
might be formed not in the remnants of H{\sc ii} region parent
clouds but ``somewhere'' along the sightlines to the latter.

The existence of dark matter in and around spiral galaxies (like
our own Milky Way) is indirectly, dynamically indicated by the
nearly flat rotation curves in the outer parts of these galaxies
that are constructed under the assumption of circular rotation of
the tracers---see e.g. Sofue \& Rubin (2001). This approximately
constant speed of galactic rotation, independent of the distance
from the galaxy centers, is usually considered to provide rigorous
proof of the presence of DM around our and other galaxies.
The analysis of stellar motions by Kuijken \& Gilmore (1989), Holmberg,
\& Flynn, (2000) provide no strong evidence for dark matter, while not ruling it out either. 
These analyzes were based on the motions of quite nearby stars. This is because the density of galactic
objects declines  outside the solar orbit, making it difficult to
construct statistically significant samples of tracers situated
outside the solar circle.
The main source of information used for construction of the
Galactic disc rotation curve is the velocities of CO clouds, which
are believed to be spatially correlated with H{\sc ii} regions,
but there have been attempts to apply other tracers. Since the
1970s, the distances used are those based on the distances of very
bright objects, such as OB stars and their clusters. The radial
velocities are measured either using H$_\alpha$ lines originating
in H{\sc ii} regions (Georgelin \& Georgelin 1976) or using
(usually) CO lines in molecular clouds (Clemens 1985). Moffat,
Jackson \& Fitzgerald (1979) tried to improve the distance
measurements using a zero-age main sequence (ZAMS) fitting method
for H{\sc ii} regions, but distances to these objects that are
normally used are those based on spectroscopic parallax. Moreover,
it is assumed that H{\sc ii} regions and molecular clouds are at
the same distances as the stars observed. This is likely to be
true in the case of H{\sc ii} regions, but these regions are
expanding shells of ionized gas and thus a great scatter in
velocities is to be expected. Interstellar molecular lines
observed in the spectra of OB stars may originate in clouds along
the sightline, but at large distances from the observed stars.

Maciel and Lago (2005) compared the Galactic rotation curve based
on a large sample of planetary nebulae with that of Brand and
Blitz (1993) based on H{\sc ii} regions. Planetary-nebulae--derived
rotation velocities are systematically lower than those for H{\sc ii}
regions. This may be either due to uncertainties of distance
measurements or due to different kinematics of H{\sc ii} regions and
planetary nebulae.

Several articles present Galaxy rotation curves based on stellar
spectrophotometry.  Liu \& Zhu (2010) used 194 carbon stars at
distances up to $R$=15 kpc from the Galactic centre and reported a
flat rotation curve with considerable scatter. A similar result
was recently reported by L\'{o}pez-Corredoira (2014), who derived
the rotation curve of the Galaxy in the range of Galactocentric
radii $R$=4--16~kpc using the proper motions of red clump giants
and near-infrared photometric data from the 2MASS survey.  A
flat rotation curve for the outer Galaxy, based on radial
velocities of red giants and horizontal branch stars was also
obtained by Xue et al. (2008) and by Bovy et al. (2012).

\begin{table*}
\caption{Data on the clusters taken from the Simbad database. Radial velocity values are rounded.
Galactic coordinates $l,b$ and the number of stars $N$
used for calculating the mean RV value are listed.}
\label{Cluster}
\begin{tabular}{lcrrcrl}
\hline
 Cluster      & l
                       & b
                              &  dist (pc)
                                       & RV$_{\sun}$ (km/s)
                                                     & N
                                                            & Ref. \\
\hline
NGC 884      &  135.1  & -3.6  & 2940  & -43$\pm$1   &  5   & Dias et al. (2002) \\
NGC 869      &  134.6  & -3.7  & 2079  & -42$\pm$2   &  54  & Kharchenko et al. (2005a) \\
ASCC 4       &  123.1  & -1.3  &  750  &  -9$\pm$9   &  4   & Kharchenko et al. (2005b) \\
Stock 2      &  133.3  & -1.7  &  303  &   2$\pm$2   &  27  & Kharchenko et al. (2005a)\\
IC 1848      &  137.2  &  0.9  & 2002  & -47$\pm$7   &  4   & Dias et al. (2002) \\
IC 1805      &  134.7  &  0.9  & 2344  & -45$\pm$11  &  6   & Dias et al. (2002) \\
\hline
\end{tabular}
\end{table*}

\begin{table}
\caption{Data on the observed stars (Northern hemisphere) and equivalent width of Ca{\sc ii} K and H lines. RV$_{\sun}$ - heliocentric radial velocity of the star.
V$_{rot}$ - calculated orbital velocity. t- Terskol, b-BOES}
\label{starsN}
\scalebox{0.6}{
\begin{tabular}{rlrrrcccr}
\hline
      Star & l     & b     & EW(K)      & EW(H)      & dist                 & RV$_{\sun}$ & V$_{rot}$ & R$_{gc}$ \\
           &       &       &    (m\AA)  &   (m\AA)   &  (pc)                    & (km/s)          &  (km/s)   &  (kpc)   \\
\hline
+56-574t   & 135.0  & -03.6 & 458$\pm$32 & 266$\pm$25 &   1700$^{+650 }_{-400 }$ &  -42  &  176   &   9.3   \\
+59-451b   &  133.4 & -1.4  & 432$\pm$18 & 294$\pm$22 &   2317$^{+308 }_{-285 }$ &  -50  &  173   &   9.7    \\
+59-456b   &  133.7 & -1.3  & 544$\pm$12 & 387$\pm$17 &   3277$^{+288 }_{-270 }$ &  -52  &  183   &  10.5   \\
+60-470t   &  133.9 & -0.1  & 453$\pm$28 & 339$\pm$27 &   3200$^{+800 }_{-600 }$ &  -44  &  197   &  10.5  \\
+60-493b   &  134.6 & 0.6   & 551$\pm$10 & 377$\pm$10 &   2976$^{+121 }_{-119 }$ &  -43  &  194   &  10.3   \\
+60-498t   &  134.6 & +1.0  & 488$\pm$26 & 330$\pm$26 &   2600$^{+1000}_{-600 }$ &  -43  &  188   &  10.9    \\
+60-499t   &  134.6 & +1.0  & 562$\pm$63 & 343$\pm$57 &   2300$^{+1200}_{-800 }$ &  -51  &  170   &   9.8    \\
+60-501t   &  134.7 & +0.9  & 525$\pm$30 & 361$\pm$21 &   2900$^{+1000}_{-600 }$ &  -43  &  193   &  10.2   \\
+60-513t   &  134.9 & 0.9   & 586$\pm$53 & 415$\pm$61 &   3478$^{+938 }_{-812 }$ &  -49  &  190   &  10.7  \\
+60-526t   &  135.5 & 0.8   & 598$\pm$34 & 390$\pm$29 &   2847$^{+285 }_{-277 }$ &  -47  &  184   &  10.2   \\
+60-594b   &  137.4 & 2.1   & 432$\pm$6  & 292$\pm$5  &   2276$^{+49  }_{-50  }$ &  -43  &  180   &   9.8  \\
+61-411t   &  133.8 & +1.2  & 492$\pm$61 & 388$\pm$68 &   4100$^{+1400}_{-2000}$ &  -57  &  186   &  11.2  \\
+61-468t   &  135.6 & +2.1  & 371$\pm$26 & 242$\pm$24 &   1800$^{+900 }_{-500 }$ &  -34  &  190   &   9.4   \\
2905t      &  120.8 & 0.1   & 247$\pm$2  & 174$\pm$3  &   1489$^{+49  }_{-49  }$ &  -27  &  214   &   8.9    \\
5394t      &  123.6 & -2.1  &  17$\pm$2.1&   8$\pm$2.1&    117$^{+15  }_{-14  }$ &  -5   &  207   &   8.1    \\
12323t     &  132.9 & -5.9  & 313$\pm$17 & 221$\pm$11 &   1878$^{+74  }_{-74  }$ &  -42  &  180   &   9.4   \\
13256b     &  132.6 & -0.6  & 608$\pm$32 & 418$\pm$23 &   3319$^{+192 }_{-192 }$ &  -52  &  184   &  10.5   \\
13267t     &  133.5 & -3.6  & 717$\pm$13 & 431$\pm$10 &   2950$^{+388 }_{-340 }$ &  -58  &  168   &  10.3   \\  
13716t     & 134.0  & -03.3 & 469$\pm$18 & 282$\pm$19 &   1850$^{+450 }_{-300 }$ &  -44  &  176   &   9.4    \\
13758t     & 134.6  & -03.3 & 483$\pm$25 & 338$\pm$27 &   2800$^{+1200}_{-700 }$ &  -40  &  197   &  10.2   \\
13831t     & 134.5  & -04.2 & 509$\pm$16 & 366$\pm$15 &   3200$^{+700 }_{-500 }$ &  -48  &  188   &  10.5   \\
13841t     & 134.4  & -03.9 & 585$\pm$11 & 363$\pm$22 &   2500$^{+400 }_{-400 }$ &  -41  &  190   &   9.9    \\
13854t     &  134.4 & -3.9  & 578$\pm$30 & 360$\pm$19 &   2467$^{+130 }_{-129 }$ &  -50  &  174   &   9.9   \\
13866t     & 134.5  & -04.2 & 530$\pm$19 & 316$\pm$11 &   2050$^{+300 }_{-200 }$ &  -49  &  170   &   9.5    \\
13890t     & 134.5  & -04.2 & 520$\pm$13 & 353$\pm$13 &   2750$^{+450 }_{-350 }$ &  -47  &  184   &  10.1   \\
13969t     & 134.5  & -03.8 & 526$\pm$14 & 333$\pm$16 &   2350$^{+400 }_{-300 }$ &  -46  &  179   &   9.8    \\
14014t     & 134.8  & -04.6 & 561$\pm$19 & 387$\pm$15 &   3100$^{+600 }_{-450 }$ &  -52  &  179   &  10.4   \\
14053t     & 134.6  & -03.9 & 581$\pm$16 & 358$\pm$15 &   2400$^{+400 }_{-300 }$ &  -43  &  185   &   9.8   \\
14134t     &  134.6 & -3.7  & 693$\pm$14 & 377$\pm$8  &   2207$^{+47  }_{-48  }$ &  -57  &  158   &   9.7    \\
14143t     &  135.0 & -4.0  & 648$\pm$8  & 340$\pm$5  &   1930$^{+31  }_{-32  }$ &  -47  &  171   &   9.5    \\
14302t     & 135.1  & -04.4 & 526$\pm$18 & 312$\pm$13 &   2000$^{+300 }_{-200 }$ &  -40  &  184   &   9.5    \\
14357t     & 135.0  & -03.9 & 543$\pm$18 & 347$\pm$17 &   2450$^{+500 }_{-350 }$ &  -51  &  172   &   9.9   \\
14434t     & 135.1  & -03.8 & 517$\pm$13 & 338$\pm$13 &   2500$^{+350 }_{-900 }$ &  -46  &  181   &   9.9    \\
14442t     & 134.2  & -1.3  & 520$\pm$35 & 375$\pm$32 &   3250$^{+1150}_{-950 }$ &  -54  &  178   &  10.5   \\
14443t     & 135.0  & -03.6 & 477$\pm$20 & 269$\pm$12 &   1650$^{+250 }_{-200 }$ &  -43  &  174   &   9.2       \\
14476t     & 135.0  & -03.5 & 507$\pm$15 & 288$\pm$17 &   1800$^{+300 }_{-300 }$ &  -45  &  172   &   9.4        \\
14818t     &  135.6 & -3.9  & 560$\pm$30 & 296$\pm$16 &   1702$^{+88  }_{-89  }$ &  -49  &  164   &   9.3        \\
14947t     &  135   & -1.7  & 451$\pm$26 & 335$\pm$23 &   3111$^{+286 }_{-281 }$ &  -61  &  163   &  10.4       \\
15558t     &  134.7 & 0.9   & 476$\pm$15 & 330$\pm$13 &   2675$^{+141 }_{-139 }$ &  -41  &  193   &  10.1       \\
15570t     &  134.8 & 0.9   & 464$\pm$19 & 350$\pm$16 &   3361$^{+188 }_{-186 }$ &  -50  &  187   &  10.6      \\
15629t     &  134.8 & +1.0  & 463$\pm$20 & 352$\pm$17 &   3450$^{+950 }_{-750 }$ &  -55  &  179   &  10.7      \\
15785b     &  135.3 & 0.2   & 537$\pm$28 & 337$\pm$19 &   2338$^{+142 }_{-140 }$ &  -45  &  180   &   9.8        \\
16310t     &  136.4 & -0.9  & 537$\pm$28 & 337$\pm$19 &   2152$^{+310 }_{-286 }$ &  -46  &  175   &   9.7        \\
16429t     &  135.7 & 1.1   & 472$\pm$17 & 307$\pm$16 &   2249$^{+169 }_{-165 }$ &  -50  &  170   &   9.7       \\
17505t     &  137.2 & +0.9  & 456$\pm$15 & 331$\pm$14 &   2900$^{+700 }_{-450 }$ &  -46  &  186   &  10.3       \\
17520t     &  137.2 & +0.9  & 494$\pm$32 & 357$\pm$29 &   3100$^{+1400}_{-850 }$ &  -48  &  184   &  10.5      \\
20336t     &  137.5 & +7.1  &  32$\pm$2  &  18$\pm$2   &   195$^{+29  }_{-38 }$  &   -3  &   210  &   8.1      \\
25940t     &  153.7 & -3    &  30$\pm$2  &  22$\pm$1.2 &   271$^{+5   }_{-6  }$  &   0.1 &  238   &   8.2         \\
236960t    &  134.6 & -1.5  & 553$\pm$46 & 427$\pm$48 &   4300$^{+1700}_{-1600}$ &  -66  &  191   &  11.4       \\
27192t     &  152.8 & 0.6   &  95$\pm$6  &  48$\pm$3.3 &   330$^{+18  }_{-19 }$  &   -2  &   237  &   8.3         \\
\hline
\end{tabular}
}
\end{table}

\begin{table}
\caption{Data on the observed stars (Galactic anticentre) and equivalent width of Ca{\sc ii} K and H lines.
All spectra are from Terskol}
\label{starsant}
\begin{tabular}{rlrrrrr}
\hline
Star        & l      & b      & EW(K)       & EW(H)       & dist                   & RV$_{\sun}$    \\
            &        &        & (m\AA)      & (m\AA)      &  (pc)                  & (km/s)            \\
\hline
43818       &  188.5 & +03.9  & 297$\pm$18  & 225$\pm$3   & 2300$^{+450 }_{-400 }$  & +16  \\
40111       &  184.0 & +00.8  & 165$\pm$3   & 99$\pm$3    &  700$^{+65 }_{-50 }$    & +12  \\
254755      &  189.1 & +03.3  & 415$\pm$19  & 315$\pm$18  & 3100$^{+1300 }_{-750 }$ & +19 \\
254699      &  188.3 & +03.7  & 317$\pm$21  & 217$\pm$13  & 1750$^{+1000 }_{-400 }$ & +16  \\
254042      &  187.6 & +03.5  & 326$\pm$21  & 247$\pm$18  & 2400$^{+1500 }_{-700 }$ & +15   \\
255055      &  188.7 & +03.9  & 301$\pm$13  & 228$\pm$14  & 2350$^{+850 }_{-650 }$  & +14  \\
255134      &  188.7 & +03.9  & 298$\pm$16  & 247$\pm$21  & 2200 phot sat           & +15   \\
251847      &  187.0 & +01.6  & 286$\pm$17  & 200$\pm$20  & 1700$^{+950 }_{-500 }$  & +15  \\
250028      &  184.9 & +00.8  & 246$\pm$11  & 188$\pm$12  & 1900$^{+900 }_{-500 }$  & +14   \\
255091      &  188.2 & +04.1  & 303$\pm$18  & 210$\pm$13  & 1750$^{+600 }_{-400 }$  & +16   \\
255312      &  188.8 & +04.1  & 282$\pm$17  & 209$\pm$18  & 1950$^{+1200 }_{-550 }$ & +13  \\
\hline
\end{tabular}
\end{table}

Discrepancies between the galactic rotation curve and a
Keplerian one are usually interpreted either in terms of
DM or MOND (Modified Newtonian Dynamics), see e.g. Milgrom (1983). 
Recent determinations of the rotation curve of M31 based on the
observations of its gaseous disk, even show a rise of
the speed of rotation in its outer parts, which cannot be understood
in terms of standard DM models or perturbations of the M31's
disk by its satellites (Chemin et al. (2009), Corbelli et al., 2010).

Quite recently, strong support for the flatness of the outer
Galactic rotation curve seems to be  provided by very precise
astrometric measurements made as part of the VERA program of
trigonometric parallaxes and proper motions of a few star-forming
regions distributed far away beyond the solar orbit: Toshihiro et
al. (2009), Reid et al. (2009);   Oh et al. (2010). These results
favor a nearly flat, or even slightly rising outward Galactic
rotation speed up to 13 kpc from the Galactic center and indicate
that this curve is similar to that of the Andromeda Galaxy.

\begin{figure}
\includegraphics[angle=270,width=9cm]{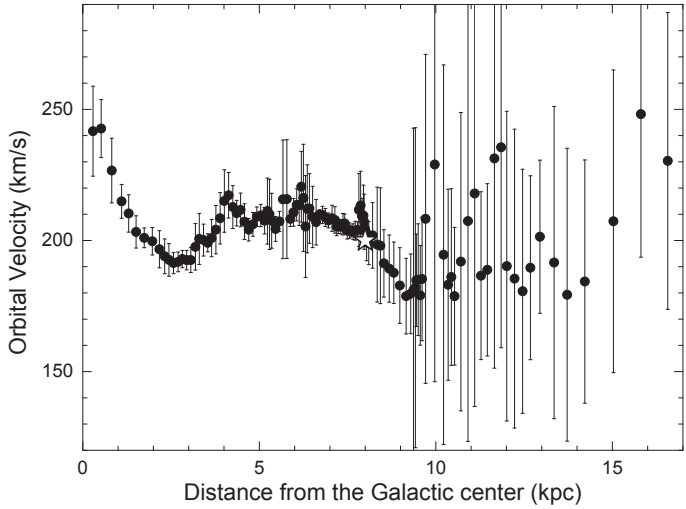}
\caption{Average galactic rotation curve based on data from the Sofue's web page.
The position of the Sun is marked as  $\bigstar$.
Outside the solar orbit there is a large scatter of orbital velocities
and measurements are scarce.}
\label{sofue}
\end{figure}

Sofue et al. (2009)  have unified the existing data on the rotation
curve of the Galaxy, presenting a single rotation curve by
re-calculating distances and velocities, adopting for the
galactocentric distance and the orbital velocity of the Sun the values
(R$_{\sun}$,V$_{\sun}$)=(8.0 kpc, 200 km/s).
The resulting curve is generally flat, but two local minima, or
dips, are prominent: at the radii 3 and 9 kpc. The 3-kpc dip is
consistent with the observed bar (or alternatively explained by a
massive ring with the density maximum at a radius of 4-kpc). The
9-kpc dip is clearly exhibited by different tracers as the most
peculiar feature in the Galactic rotation curve. The authors
explain it by a massive ring with the density peak at a radius of
11-kpc. This great ring may be related to the Perseus arm.  It is
evident that the sample of tracers situated outside the solar
orbit (and believed to be intrinsically related to the Galaxy) is
much smaller than that of inner tracers. Moreover, the scatter of
individual determinations of distances and radial velocities grows
in a stepwise manner outside the solar orbit (see Fig.~\ref{sofue}).
These uncertainties in the measured radial velocities
look strange; they very likely are from measurements of very
broad H$_{\alpha}$ lines originating in H{\sc ii} regions, which do not
allow precise radial velocity determinations.

The sudden growth in the scatter of points situated outside the
solar orbit, seen in Fig. \ref{sofue}, follows the use of
different methods to determine distances and orbital velocities
inside and outside the solar orbit. Inside the solar circle, the
method of tangent points is typically used. The orbital velocity
of the point (tangent point) is uniquely determined by the maximal
radial velocity in a given direction. At this point the radius
vector of the observed object $R_{GC}$ is perpendicular to the
sightline. This gives a right triangle, i.e. all the triangle's
sides and angles are determined---including the galactocentric
distance of the object observed at the tangent point. The result
is scaled to the assumed radius of the solar orbit. The method of
tangent points cannot be applied outside the solar circle. Thus,
the rotation curve of galactic peripherals depends critically on
measurements of individual distances and radial velocities.
Such methods are called ``clear'' if they apply direct distance
and radial velocity measurements; otherwise models based on
certain assumptions are applied to large samples of objects.

\section{Distance determination problems and why Ca{\sc ii} is better tracer}

The resulting, generally flat, rotation curve is an average one.
The same concerns the rotation curves of other galaxies. The
distances to tracers, especially for the outer part of the Galaxy,
are either based on angular sizes of H{\sc i} clouds in the
Galactic disk (Merrifield, 1992), or on spectroscopic parallaxes
of OB stars (plus ZAMS fitting). The former require assumptions of
their linear sizes and thus remain very uncertain. The latter
suffer large errors because of the scarcity of OB stars within the
range of detectable trigonometric parallaxes, making the
calibrated M$_V'$s uncertain. Moreover, massive OB stars are
frequently spectroscopic binaries or, in general, variables. The
apparent and absolute magnitude in photometric equations should be
measured and calibrated at the same phase. Such a calibration,
i.e. for different phases of variability, has not been done for
bright, OB stars. The estimated distances, both those based on
linear sizes of H{\sc i} clouds and those resulting from ZAMS
fitting thus suffer large errors, which themselves are difficult
to estimate. A clear example will be shown later. However, the
very error bars of the data, around the average rotation curve
beyond the solar circle (Fig. \ref{sofue}) is shown only for the
orbital velocities! The errors of distances are not given in the
homepage of Sofue \footnote{\url{http://www.ioa.s.u-tokyo.ac.jp/~sofue/}}.

The distance errors have not been estimated. The observed
scatter should indicate not only significant errors in the
distance and radial velocity determinations of the tracers used,
but also more complex intrinsic kinematics of the considered outer
disk. Visual inspection of images of spiral galaxies seen face-on
leads to the conclusion that while the inner parts of the discs
show very regular spiral structure, the latter is heavily
perturbed in the outer parts of discs.

Inspection of pictures of external spiral galaxies seen edge-on
reveals well-distinguished, narrow, dust--gaseous thin discs extended
up to the visible borders of these galaxies. Such a thin disc
structure is apparently perpendicular to the galactic rotation
axis. On the other hand, the  stellar (especially evolved) component forms  the thick disk
in our Galaxy (and in other spiral galaxies), apparently
influenced by other forces than just the gravity of the central
bulge and the centrifugal force. Thus its kinematics is much more
complex than that of the thin disc. This may be a result of a
strong disturbing influence of possible merging effects. To
properly identify these sources we first need  a clear picture of
the true, ``basic'' kinematics of the original thin disc, e.g. its
rotation curve. We should expect to obtain this for our Galaxy from the
observed motion of the most representative and reliable thin disc tracers.
The latter are the interstellar clouds and OB stars recently formed out of this diffuse matter.

In this work, we introduce a new method for estimating
the Galaxy rotation curve based on measurements of intensities
and radial velocities of interstellar Ca{\sc ii} lines in the
optical wavelength range, with just a single assumption: circular
orbital motion of interstellar, optically thin clouds.

The presence of DM inside and/or around our Galaxy was recently
questioned by Moni Bidin et al. (2012). Estimating the dynamical
surface mass density at the solar position between $Z$ = 1.5 and 4
kpc from the Galactic plane, the authors concluded that the local
density of DM is at least an order of magnitude below the standard
expectations. However,  Bovy \& Tremaine (2012) used the model
with different assumptions (although simplified) and concluded the
accordance of observations with the standard DM paradigm. This
fact makes it very important to carry out other observational
tests of the existence of DM; that the same data interpreted using
different models leads to contradictory conclusions proves that
that observational data are insufficient.

Our recent papers (Megier et al. 2005, 2009) demonstrated two
important facts. First: interstellar space in the thin disk of
the Galaxy is rather evenly filled with optically thin clouds,
revealed by the Ca{\sc ii} H and K lines. The same Doppler
components that reveal the presence of many clouds along a
sightline can be found in Na{\sc i} and K{\sc i} strong interstellar lines.
Second: the column density of Ca{\sc ii}
and, in practice, the equivalent widths (EWs) of
these lines since saturation effects are low; (the
latter are much stronger in Na{\sc i} lines) can
be used to infer distances to
the observed stars. With growing distance from the Sun (in the
outer part of the Galaxy) the observed Doppler components are more
and more Doppler shifted along any sightline---probably due to the
differential rotation of the Galaxy.

\begin{figure}
\includegraphics[angle=270,width=9cm]{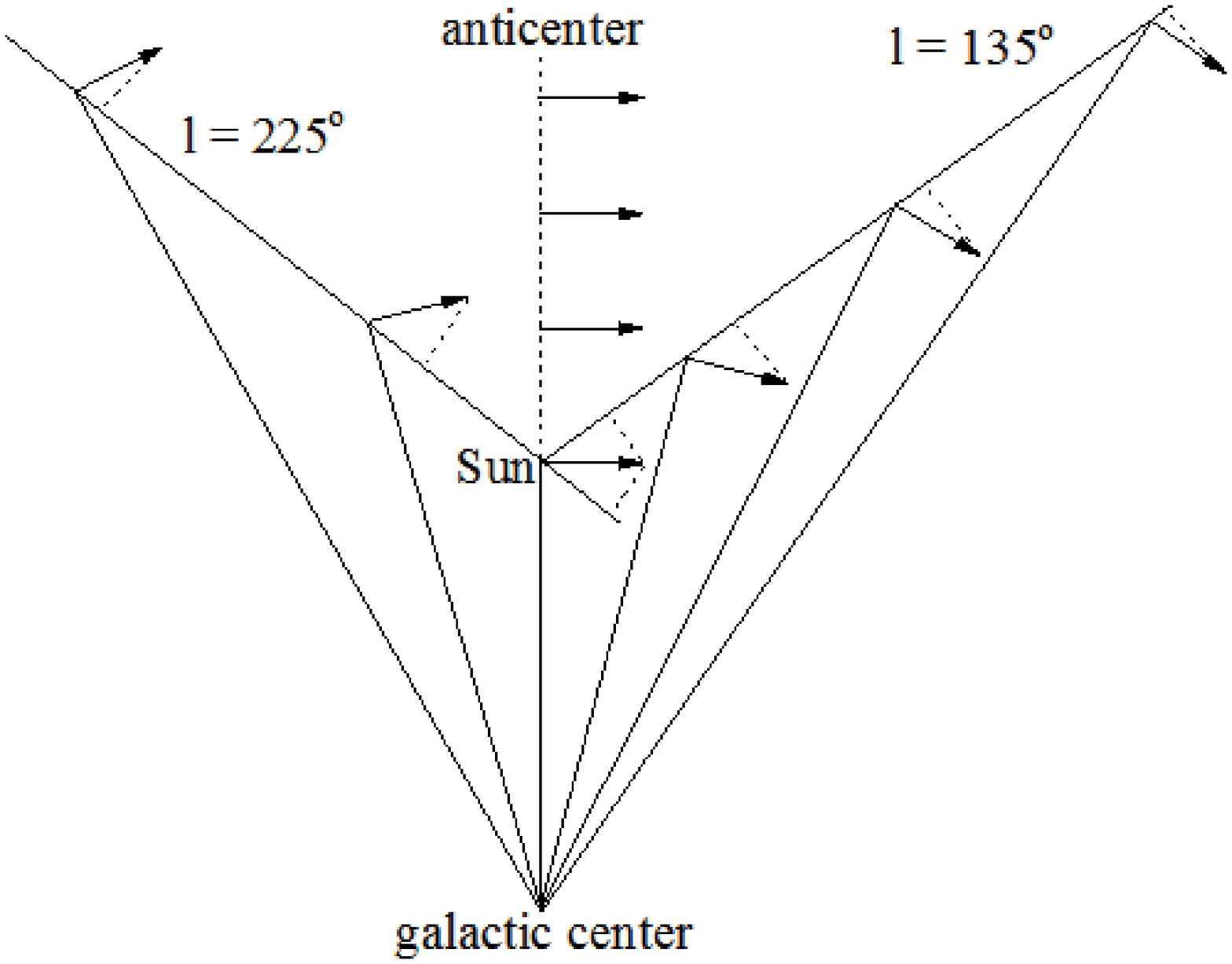}
\caption{Schematic diagram showing that radial velocities towards
the observer should increase with distance at $l=225${\degr} or
decrease with distance at $l=135${\degr}. Those observed towards
the galactic anticenter should be close to zero if the rotation
curve is flat and the orbits are circular. This diagram of the
Galaxy is not made to scale; the aim is to explain the effects
expected.} \label{scheme}
\end{figure}

The direction towards $l=135${\degr} looks especially interesting.
Trigonometry, as shown in Fig.~\ref{scheme}, shows that the blue-shifts of Doppler
components grow with distance in this
direction. This is important because we have good observations of
Doppler splitting, even with medium resolution, and because the
broadening of the lines, caused by the Doppler splitting, prevents
them from strong saturation. One should expect that the Doppler
shifts grow with distance and thus the most blue-shifted Doppler
components (in the $l=135${\degr} direction) are formed in the most
distant clouds (assuming that a sightline intersects a couple of
clouds at different distances). It is interesting to check whether
the same Doppler structure can be found in other interstellar
features and whether the observed stars, which are the most
distant objects along a sightline) share the radial velocity with
the most distant clouds.

\begin{figure}
\includegraphics[angle=0,width=9cm]{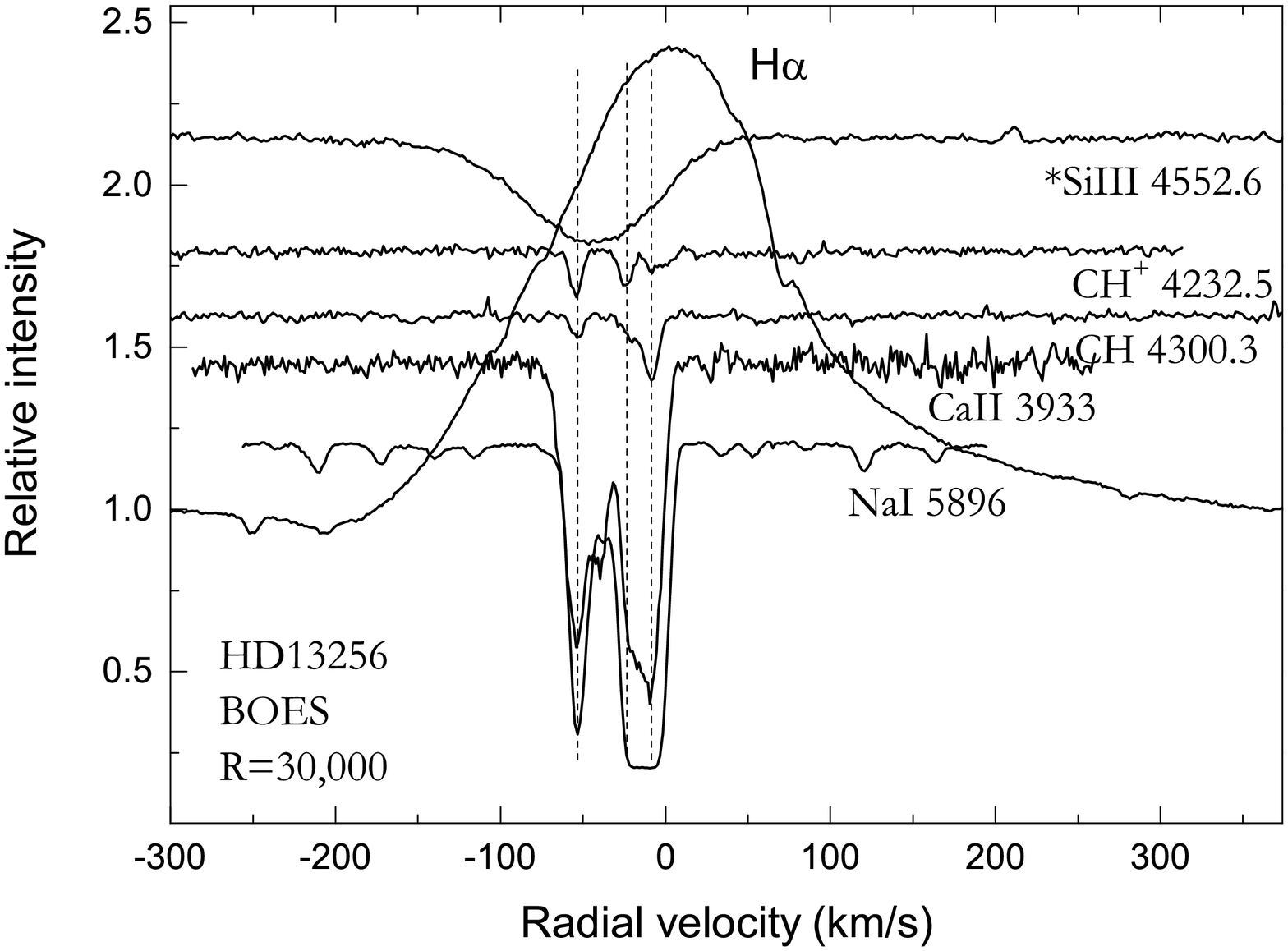}
\caption{Radial velocity components seen in different
interstellar features toward HD13256. It is important that the
main CH component does not share the radial velocity of the
stellar line.
The strongest $CH^+$ component is apparently formed near the star, while the strongest
components of neutrals are formed far away from the star but much closer to
the Sun.}
\label{radvels}
\end{figure}

Figure~\ref{radvels} clearly shows that in any of the interstellar
features one can see the same set of Doppler components.
Apparently the Doppler shifts grow with distance, which makes
possible distance estimates to individual clouds. One can observe
two important facts: intensity ratios of the same radial velocity
components of features, likely to originate in the same
environments, are different; radial velocities of the observed
stars may not be related to any interstellar component. The stars
we are interested in are likely to be members of clusters or OB
associations; many of them belong to binary systems. Thus, these
stars are likely to orbit around local gravity centers, while
interstellar clouds, situated far away from any such centers, only
show orbital motion (around the Galaxy center) and, along the
chosen sightlines, the radial components of  orbital motion only
are seen. Thus interstellar clouds are much better tracers of the
radial velocity components of orbital motion than stars. In the
$l=135\degr$ sightline, the farther a cloud is, the more (in
general) it is blue-shifted.

\begin{table*}
\caption{Column densities of CH and CO interstellar molecules. YS
-- Sheffer et al. (2007),  b -- Burgh et al. (2007), S -- Sonnentrucker et al. (2007), Lyu -- Lyu et al. (1994). CH -- our measurements. }
\label{relcoch}
\begin{tabular}{lrrrr}
\hline
Star        & N(CH) & error  & N(CO) & error    \\
            & $\times$10$^{12}$    &        & $\times$10$^{15}$   &          \\
\hline
HD22951YS   & 21.36 & 1.56  & 0.18   & 0.01     \\
HD23180YS   & 12.54 & 1.24  & 0.68   & 0.01     \\
HD24398YS   & 23.52 & 2.91  & 1.79   & 0.05     \\
HD24534b    & 22.3  & 2.61  & 13.49  & 6.5      \\
HD27778b    & 38.05 & 3.22  & 11.22  & 3.8      \\
HD73882S    & 34.02 & 1.7   & 35.5   & 17       \\
HD91824b    & 6.75  & 1.35  & 0.04   & ---      \\
HD99872YS   & 13.35 & 0.03  & 0.45   & 0.04     \\
HD147888S   & 21.84 & 0.04  & 2      & 0.38     \\
HD149757Lyu & 25.66 & 0.82  & 1.58   & ---      \\
HD154368YS  & 60.54 & 1.07  & 2.67   & 0.55     \\
HD163758b   & 7.62  & 0.35  & 0.03   & ---      \\
HD203374YS  & 23.34 & 0.28  & 2.55   & 0.43     \\
HD206267b   & 26.4  & 4.71  & 12.9   & 4.5      \\
HD207198b   & 45.21 & 5.62  & 3.4    & 1.7      \\
HD207538YS  & 44.34 & 4.07  & 2.34   & 0.24     \\
HD210121S   & 30    & 2.73  & 6.76   & 1.25     \\
HD210839b   & 33.43 & 1.86  & 2.6    & 0.5      \\
HD303308b   & 10.64 & 0.65  & 0.04   & ---
   \\
\hline
\end{tabular}
\end{table*}

\begin{figure}
\includegraphics[angle=0,width=8cm]{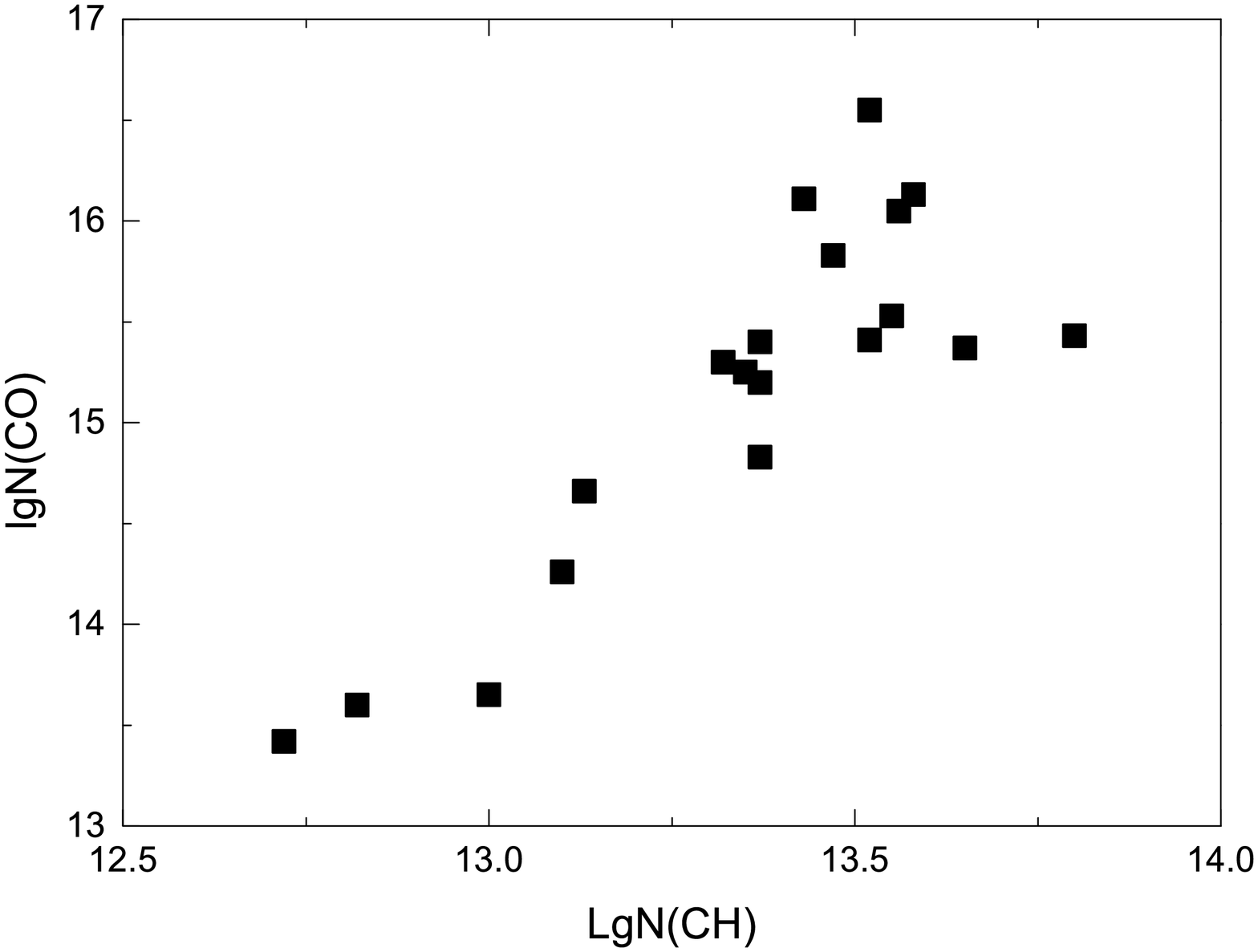}
\caption{Correlation between column densities of CO and CH
molecules
for the stars in Table~\protect\ref{relcoch}.
A positive correlation is obvious.} \label{relcoh}
\end{figure}

The published rotation curve of our Galaxy has in many cases been
based on radial velocities of CO radio lines (Clemens 1985). In
these cases, distances were estimated to the observed stars. It
seemed natural that molecular lines originated in relic clouds
from which the OB stars recently formed. It seems quite natural
that all interstellar molecules are spatially correlated. If the
latter is true one can expect some intensity correlation between
column densities of any two interstellar molecules (except,
perhaps CH$^+$). We have compiled CO column densities from several
publications in Table \ref{relcoch}. We measured column densities
of the easily accessible CH radical for these targets.
Figure~\ref{relcoh} demonstrates a reasonably tight correlation
between the two species. This favors the hypothesis that the two
molecules are spatially correlated. Thus their radial velocities
can reasonably be expected to be identical.

Figure~\ref{radvels} clearly demonstrates that the main Doppler
component of the CH 4300~\AA\ line is of a different velocity to
the Si{\sc iii} stellar line shown (and, consequently, is formed
in a cloud situated much closer to the Sun). However, the weaker
CH components are still visible as far towards the violet as those
of the Ca{\sc ii} lines. These severely shifted components of
Ca{\sc ii} and CH are of the same radial velocity and thus they
clearly can only be radial components of the orbital motion of the
most distant cloud. On the other hand the stellar line contains
some additional velocity, likely to be caused by motion around the
center of a local cluster or association. It is also evident that
the distance to the strongest CH component is much less than that
to either the star or the most distant cloud. Apparently
translucent interstellar clouds, moving only around the galactic
center, are the best tracers of the structure of the thin disc and
of the kinematics of the latter.

\begin{figure}
\includegraphics[angle=270,width=9cm]{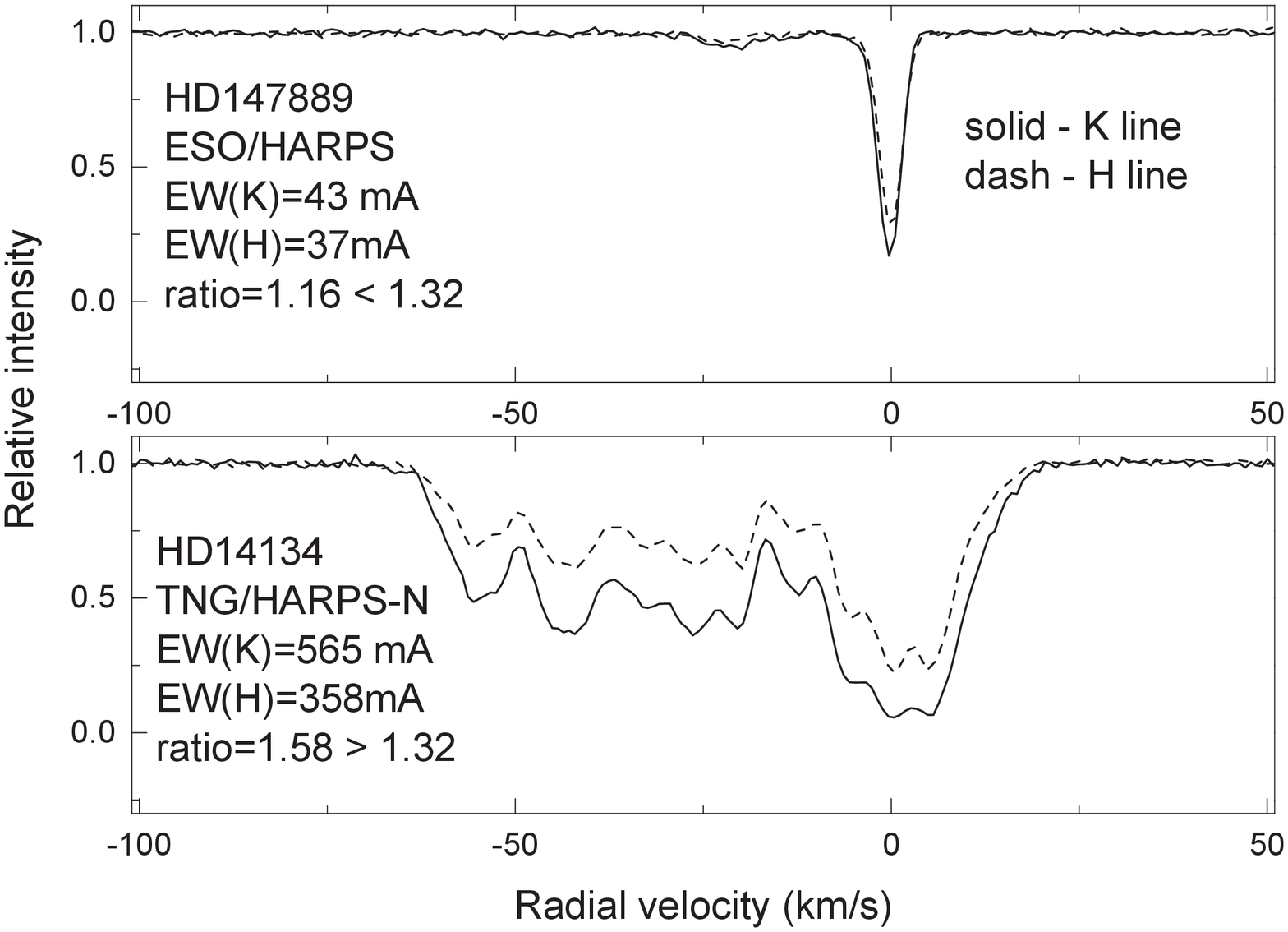}
\caption{Evidence that the clouds revealed by Ca{\sc ii} lines
observed towards the direction $l=135${\degr} evenly fill the
galactic disk  (number of components grows with distance, each
new one is more and more blue shifted) and thus, that the H and K
lines remain free of strong saturation while saturation may be
evident towards nearby objects.} \label{idea}
\end{figure}

\begin{figure}
\includegraphics[angle=270,width=9cm]{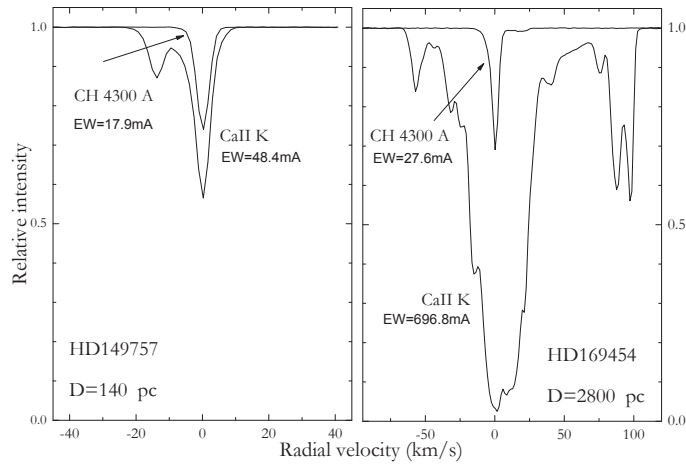}
\caption{Example of how distances grow with
Ca{\sc ii} line intensity. The CH radical is apparently abundant only in
relatively few clouds along the long sightline, making it a
poor tracer of galactic rotation.}
\label{distance}
\end{figure}

Translucent interstellar clouds are reasonably evenly distributed
in interstellar space. Ionized Ca clouds seem to be good
representatives of the gaseous thin galactic disc, supposedly
better than any other of its constituents:  Na{\sc i} lines are
usually completely saturated, making them useless, K{\sc i} lines
are in many cases too weak to allow all the components to be
traced; the same latter problem also occurs for molecular
features. Objects which appear to be unevenly distributed are
interstellar clouds, revealed by molecular lines (Fig.
\ref{distance}). As seen from Fig.~\ref{radvels}, molecular lines
are very likely to be formed in all translucent clouds, but many
components are very weak and thus fall below the detection limit.
It is commonly accepted that OB stars are spatially related to
molecular clouds out of which they have recently been formed;
however, such remnant clouds are not necessarily situated along
our sightlines. If an interstellar cloud shares the radial
velocity with the star it may be such a relic object. If the
radial velocity is drastically different, then the cloud is likely
to be far away from the observed star; see Fig. \ref{radvels}.
Thus, it is very risky to use distances calculated for OB stars
and radial velocities of the main (strongest) components of
molecular lines. Apparently the Ca{\sc ii} features originate in
many individual clouds, thin enough to get very strong lines that
are not heavily saturated (Fig.~\ref{idea}). On the other hand,
the main components of CH or CN spectral bands may be strong in
the spectra of nearby objects and very weak in distant ones (Fig.
\ref{distance}, Fig. \ref{radvels}).

The observed scatter in the published rotation curves may indicate
not only significant errors in the distance (and uncertainties in
the radial velocity) determinations of the tracers used, but also
more complex intrinsic kinematics of the outer disc. Moreover, in
all the above mentioned rotation curve determinations, little
attention has been paid to the question of the representativeness
of the (most distant) used tracers of the galactic disc. The most
distant tracers, used to do measurements depicted in
Fig.~\ref{sofue}, may be non-members of the Galaxy. Their
rotational velocities are very large and the distances are equal
to that of the LMC. We expect that thin disc objects, OB stars and
gas clouds, to be the most reliable members of the thin galactic
disk. The tiny clouds revealed by the H and K lines fill the disc
almost continuously (Fig. \ref{idea}) and thus all of them
dynamically belong to the Galaxy beyond any doubt because the
radial velocities grow continuously up to the observed star --
apparently formed recently inside the thin galactic disc.

The main purpose of the present paper is to delineate and analyze
the rotation curve of the outer part of the gaseous disk of the
Galaxy, over the distance of a few (up to 3--4) kiloparsecs beyond
the solar circle, where the flat or Keplerian character of the
rotation curve should be clearly distinguishable. As stated above,
this part of the published galactic rotation curves suffers large
uncertainties because of insufficient statistics of the tracers
observed, and because of large scatter seen in the individual
measurements of distances and radial velocities. On the other
hand, all the published Galactic rotation curves show a distinct
dip in this region and the spatial localization of this dip is
about 1--2~kpc beyond the radius where the observed density of the
visible disk starts to decline (drop) outwards, which may suggest
that these indicate the border of the Galactic Disk.

To construct the rotation curve over the selected regions we
propose to use interstellar gas clouds revealed by Ca{\sc ii}
spectral lines. This choice is based on at least two reasons:
\begin{itemize}
\item
recently we have demonstrated (Megier et al. 2005, 2009) that the
equivalent widths of interstellar Ca{\sc ii} lines, observed in
the thin Galactic disk, correlate tightly with distance.
\item
the relatively low level of saturation of even very strong
Ca{\sc ii} (in the sense of equivalent width) lines, seen in the near UV,
clearly proves that the whole space in the Galactic disc is
quite evenly filled with rather tiny (optically thin) clouds,
producing (almost) unsaturated components, more and more Doppler
shifted along a chosen sightline---probably due to the differential
rotation of the Galaxy (see Fig. \ref{idea}).
\end{itemize}
These two reasons make Ca{\sc ii} clouds the
best tracers of the thin disk structure.

Let us assume that the equivalent widths of the near UV
Ca{\sc ii} lines increase with distance by adding subsequent
optically thin clouds---thin enough to produce unsaturated H and K
lines. In general these clouds participate in the differential
rotation of the Galaxy. Depending on a chosen direction, if they
move on circular orbits, radial velocities of more and more
distant clouds should be more and more blue- or red-shifted (Fig.
\ref{idea}, Fig. \ref{scheme}). Thus the most shifted H and K
Doppler components (which share radial velocities with all other
interstellar absorptions but are stronger and thus best seen)
should measure the radial velocity of the most distant cloud which
should be very close to that of the observed star: the farthest
object along the sightline (Fig. \ref{lines_agr}). The total
column density should measure the distance to the observed star,
or more precisely, the farthest Ca{\sc ii} cloud along the line
of sight (Fig. \ref{radvels}). The formula derived by Megier et
al. (2009) from 262 sightlines is:

\begin{equation}
  D_{Ca{\sc ii}} \mathrm{[pc]} = 77 +
  \left( 2.78 +
  \frac{2.60}{\frac{\mathrm{EW(K)}}{\mathrm{EW(H)}} - 0.932}
  \right) \mathrm{EW(H)},
  \label{e-CaII-distance}
\end{equation}

where EW(H) is measured in m\AA. This formula can be applied if
EW(K)/EW(H) $>$ 1.32, which guarantees that the saturation level
is low enough. Each pair of measurements: distance (from Ca{\sc
ii} column density) and radial velocity (from Ca{\sc ii} line
profiles) can be done from the same spectrum. This strongly
reduces the number of observations needed to construct the
rotation curve and removes the ambiguity that follows from unknown
velocity components in stellar spectra. Let's emphasize that
stellar lines observed in spectra of hot, OB stars are broad which
makes difficult to resolve individual Doppler components. On the
other hand our method allows the selection of well-defined Doppler
components of Ca{\sc ii} lines to measure the radial velocity of
the most distant clouds along the chosen sightlines. Apparently,
the main components of other interstellar lines, in particular,
molecular ones, are likely to originate in clouds situated much
closer than the observed OB stars---see Fig. \ref{idea} and Fig.
\ref{radvels}.

As it was shown by Megier et al. (2009, see their Fig. 6),
the Ca{\sc ii}-method exhibits a good agreement with distances of
OB-associations, i.e. Eq. \ref{e-CaII-distance} should give
reasonable distance estimates up to $\sim$3 kpc in the Galactic
plane. In the current study we slightly exceed this value, though
most of our targets are closer than 3 kpc. The good agreement of
Ca{\sc ii} distances and those based on OB-associations (Megier et al.
2009, see their Fig. 6) is not the only proof of the correctness
of Ca{\sc ii} distances. Fig. 5 proves that the Ca{\sc ii} clouds are
distributed quite evenly, at least in the l=135 direction. The
distance to HD 14134 is 2.2 Kpc according to both Ca{\sc ii}and
spectroscopic parallax methods.

\begin{figure}
\includegraphics[angle=270,width=9cm]{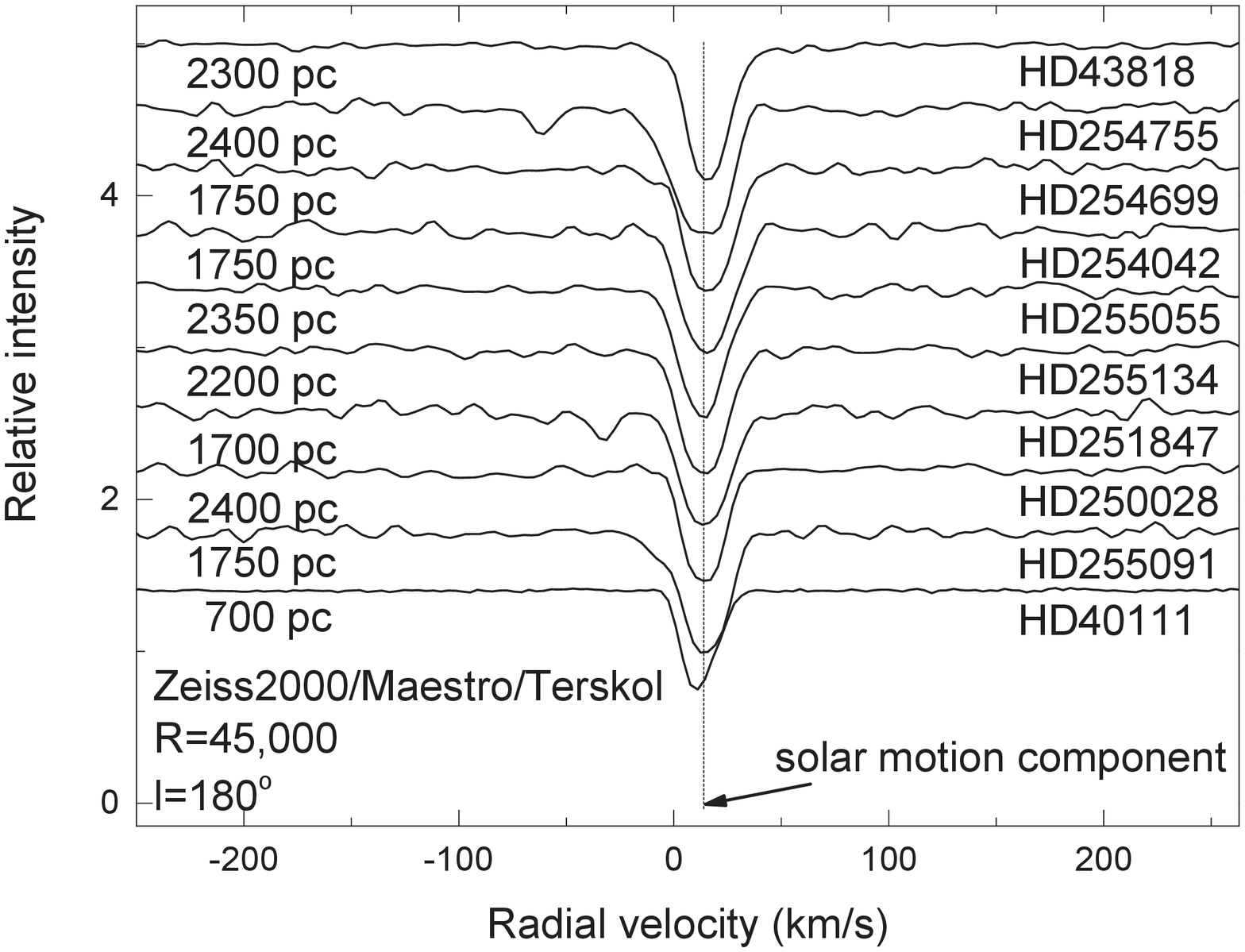}
\caption{Ca{\sc ii} lines observed towards the galactic
anticentre. There is no obvious Doppler splitting. The dashed
vertical line represents the solar motion  radial component
relative to the LSR. The orbits of presented Ca{\sc ii}
clouds seems are circular!} \label{circle}
\end{figure}

\begin{figure}
\includegraphics[angle=270, width=9cm]{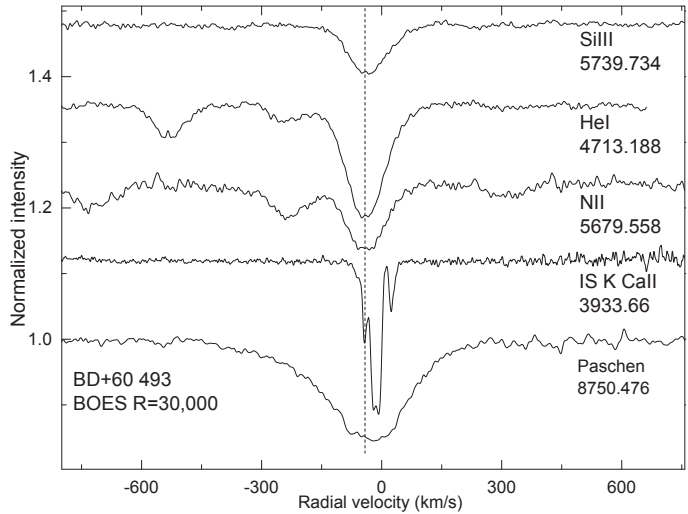}
\caption{Radial velocities of interstellar clouds growing up to
the limit of the stellar radial velocity in the direction to
$l=135${\degr}. The interstellar clouds are situated between
the observer and the star, i.e. the star is, among the depicted objects,
the most distant one and so its radial velocity (negative)
should be the greatest in that direction along this sightline.
\label{lines_agr}
}
\end{figure}

How valid is our assumption that our Ca{\sc ii} absorbers
rotate on circular orbits? Circular orbits are usually assumed,
but our method allows this assumption to be checked observationally.
The stars observed towards the galactic
anticentre are listed in Table \ref{starsant}.
Figure~\ref{circle} shows
the profiles and radial velocities of the Ca{\sc ii}
K line, seen in the spectra of our targets.
The assumption of circular orbits is clearly a good approximation for
this choice of tracers.

Neither binarity of the chosen targets (clouds) nor pulsations
affect the observed radial Ca{\sc ii} velocities (Fig.
\ref{binarity}). Also, uncertainty of determination of the profile
center is much lower in the case of interstellar absorption lines
than for the stellar spectra of fast rotators. With a resolution
of R$>$30,000, the former is better than $\sim$1.3 km/s. The
distance estimates only require one calibration: distance vs.
intensities of Ca{\sc ii} lines---rather than requiring several
calibrators, as is the case for spectroscopic parallax. This
provides another argument in favor of using the gaseous component
of the thin disk to determine the rotation curve of the Galaxy.
This has not been done previously, although, quite recently,
Bobylev \& Bajkova (2011) used the Ca{\sc ii} distances from our
work by Megier et al. (2009) to determine kinematical properties
of the local part of the Galactic disc. We stress the point that
among interstellar features seen in stellar spectra, only
interstellar Ca{\sc ii} lines show a reasonable choice of Doppler
components, correlating well with distance, in contrast to other
interstellar atoms/molecules. CH lines and other similar features
may trace the dust lanes in spiral arms, one of which is quite
close to the Sun---about 1 kpc. Since the Ca{\sc ii} clouds are
ubiquitous, filling the galactic disc almost uniformly, they form
the dominating gaseous thin disc component of the Galaxy (Fig.
\ref{idea}).

The selection of a proper tracer to establish the Galactic
rotation curve is crucial. Until now radial velocities have been
measured either for H$_\alpha$ lines of H{\sc ii} regions or for
molecular features. Distances are usually those of the observed OB
stars, sometimes determined by the ZAMS fit. Figure~\ref{radvels}
demonstrates the ambiguity of the choice made. It is natural that
the radial velocity of H$_\alpha$ is very different from that of
any component of the CH molecule.

Moreover, the radial velocity of the stellar Si{\sc iii}
4553~\AA\ line is close to one of the components of CH.
We can choose the most blue-shifted component of any of the
interstellar lines as the farthest one. The even distribution of
Ca{\sc ii} clouds makes the choice most natural. Thus only the
tiny clouds revealed by the Ca{\sc ii} lines allow to reasonably
relate distances and radial velocities; the latter measured with a
very high precision of the order of 1 km/s. The kinematics
(rotation) of this component seem to contain a key for checking
for the possible presence of local DM in the Milky Way. The
simplest way to find support for the existence of local DM is to
compare the observed rotation curve, based on Ca{\sc ii} tracers,
with e.g. the ``Keplerian'' curve determined by the distribution
of the visible matter only and the ``flat'' curve.  In
opposite to many ``model'' recent works (e.g. Bovy et al. (2012),
Xue et al. (2008)) our attempt may belong to the ``clean''
category in the sense it is based on direct measurements of
individual objects; instead of being an attempt to fit a model
which follows a set of assumptions. This is why we do not
construct any model, but just assume either Keplerian or flat
rotation curve.

To construct a kinematical map of the outer part of the Galaxy one
needs reliable Galactic disc tracers with well-determined
distances and radial velocities. Uncertainty in determining the
rotation curve in the outer part of the Galaxy follows the
uncertainties of estimates both of these.

In the case of H{\sc i} clouds, the distances are inferred from
their angular sizes; this requires assumptions about their linear
sizes (which are very uncertain). The H{\sc ii} regions should be
at the same distances as those of their central OB stars. However,
the distances to OB stars, based on spectroscopic parallax, are
uncertain for several reasons:
\begin{itemize}
\item
calibrations of absolute magnitudes for
a given spectral type and luminosity (Sp/L) are uncertain
because of poor statistics of OB stars that have reliably
measured trigonometric
parallaxes
\item
all stellar parameters (in particular $m$ and $M$) should be used at
the same phase of possible variability; calibrations of $M$ for
different variability phases do not exist
\item
calibrations of intrinsic $B-V$ colors for a given Sp/L (the
latter frequently seriously flawed); these are less important but
non-negligible
\item
determinations of extinction curves and total-to-selective
extinction ratios which may vary from object to object; using a mean
extinction law may lead to very serious errors---on average
growing with distance but not possible to be properly estimated
\item
the unsolved problem of neutral (grey) extinction.
\end{itemize}

\begin{figure}
\includegraphics[angle=270, width=9cm]{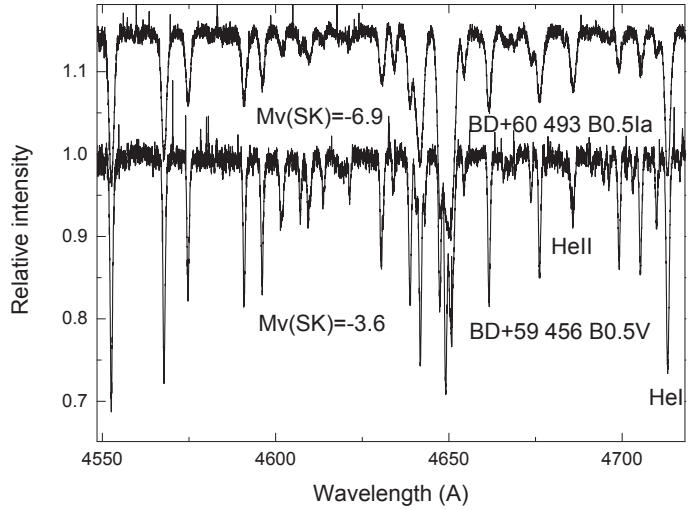}
\caption{Spectral classification errors leading to huge errors
in distances based on the spectroscopic parallax method.
\label{klasyk} }
\end{figure}

\begin{figure}
  \includegraphics[angle=270,width=9cm]{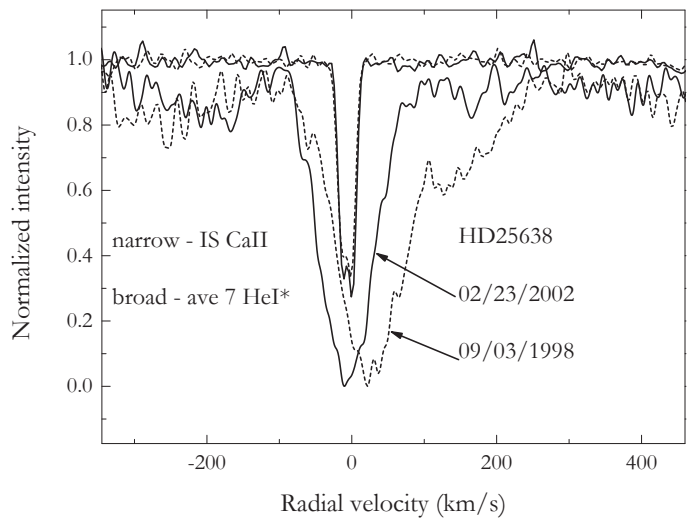}
  \caption{Illustration of how binarity or pulsations of the
    stellar object can influence the measured radial velocity. The radial
    velocity measured using Ca{\sc ii} lines remains unchanged.}
  \label{binarity}
\end{figure}

To illustrate this, let us consider two stars from our list (Table
\ref{starsN}): BD +59-456 and BD +60-493. According to the Simbad
database their Sp/L's are B0.5V and B0.5Ia, respectively. For
distant objects, which are needed for determining the rotation
curve, we cannot expect many independent classifications in the
literature. Using the spectroscopic parallax method we estimated
the distances to the above objects as 1600 and 2800 pc
respectively (assuming R$_V$=3.1). Fig. \ref{klasyk} demonstrates
our high resolution spectra of both targets. It is evident that
the spectral type is the same (note the identical ratio of
HeI/HeII lines) but the luminosity classes stated in Simbad are
evidently incorrect. Indeed, the spectral lines in the ``dwarf''
are narrower than in the ``supergiant''. If BD+60-493 has been
classified correctly, then BD+59-456 should be of the luminosity
class Ia-0 rather than V and thus, according to Schmidt--Kaler
(1982), be of the absolute magnitude $M$=-8.2. This would double
the spectroscopic distance. In the Schmidt--Kaler (SK) system,
there is no intermediate Sp/L allowing for an interpolation
between an absolute magnitude of $-6.9$ (for Ia) and or $-8.2$)
(for Ia-0). If our method, based on Ca{\sc ii} lines, is
applied, the distances are 3300 and 3000~pc, respectively, which
is very reasonable.

In practice, all these uncertainties add one to another and the
final result (distance) appears to be very questionable. It is not
possible to reliably estimate errors in distance estimates made
using spectroscopic parallax. Our method only involves the total
intensities of Ca{\sc ii} lines (distance---Fig.~\ref{distance})
and the positions of the most blue-shifted Doppler components
(radial velocity---Fig.~\ref{radvels}, Fig.~\ref{lines_agr}).
These do not need any calibration except for the observed relation
of Ca{\sc ii} column density and parallax (Megier et al. 2009).
Our measurements are reasonably precise until the Ca{\sc ii} lines
become heavily saturated (EW(K)/EW(H) $>$1.32). Moreover, any
distance--radial-velocity pair can be reliably determined from a
single spectrum. Stellar binarity, variability and possible
motions around local gravity centers (e.g. clusters) do not
influence either distance or radial velocity measurements (Fig.
\ref{binarity}). The rotation curve may be constructed from a
homogeneously measured set of distances and radial velocities for
interstellar Ca{\sc ii} clouds, which fill the galactic disc much
more evenly than stars and clusters, being for the same reason the
dominant gas component of the thin Galactic Disc.

Here, we propose a new tracer of the Milky Way rotation curve:
Ca{\sc ii} clouds, possessing a variety of advantages in
comparison to ``classic'' tracers, mainly because of requiring
fewer assumptions, thus reducing sources of systematic error.

\section{\protect Method}

Assuming circular orbits of objects (stars and clouds) in the
galactic disc we have selected two directions (galactic longitudes
180{\degr} and 135{\degr}) which enable checking galactic rotation
according to Fig. \ref{scheme}. In Fig. \ref{scheme} it is shown
how the observed radial velocity should decrease with increasing
distance from the Sun in the 135{\degr} direction. In the
direction to the Galactic anticentre the observed radial velocity,
related to the local standard of rest (LSR), of every object is
expected to be close to zero (if it is true that the orbits of
thin disc objects are circular).

The Sun does not move on an exactly circular orbit. The total
velocity of our star, relative to the LSR, defined by young O--B5
stars and B8--A0 supergiants in the distance range 0.2--3.0~kpc,
derived using the Skymap Catalogue (Zhu, 2006) distance scale, is $V_{\mathrm s}$ =
20.1$\pm$0.4km/s towards ${l_{\sun}}$ = 51{\degr}.2$\pm$1{\degr}.2
and ${b_{\sun}}$ = +22{\degr}.9$\pm$1{\degr}.1. Using the
Hipparcos distance scale the above values are respectively:
$V_{\mathrm s}$ = 16.9$\pm$0.4~km/s, ${l_\sun}$ =
47{\degr}.2$\pm$1{\degr}.3 and ${b_\sun}$ =
+22{\degr}.9$\pm$1{\degr}.2 (Zhu 2006). Miyamoto \& Zhu (1998)
previously determined somewhat different values: $V_{\mathrm s}$ =
19.1$\pm$0.5~km/s, ${l_\sun}$ = 49{\degr}.2$\pm$1{\degr}.6 and
${b_\sun}$ = +21{\degr}.9$\pm$1{\degr}.3. In this paper we have
adopted the average values $V_{\mathrm s}$ = 18.5~km/s, ${l_\sun}$
= 49{\degr}.2 and ${b_\sun}$ = +22{\degr}.9; our model rotation
curves are calculated using the latter averages. The solar motion
with respect to the LSR is especially useful for our method, since
it is nearly perpendicular to the direction $l=135${\degr}, i.e.
it has a very weak influence on our measured radial velocities.
This is why the radial velocity is close to zero for nearby
objects at $l=135${\degr} (see Fig. \ref{north-new}).

 On the other hand our estimated value of the Sun's radial
motion, derived from radial velocities of Ca{\sc ii} clouds in the
direction l=180 is close to 12km/s (Fig. 7) quite accordant to
that, given by Bovy et al. (2012).

Our proposed construction of the galactic rotation curve is based
on distance estimates and radial velocities of the most
blue-shifted Doppler components of the ionized calcium
(Ca{\sc ii}) lines in the direction $l=135${\degr}. In the
direction of the anticentre, the expected radial velocities are
zero: in the current study we check this as a control on the
validity of our method.

The rotation curve is a smoothed average of the velocity field; it
is the relation that gives the velocity of rotation on circular
orbits with respect to the center of a galaxy as a function of
distance from its center. However, various deviations from
circular movement, e.g. peculiar movements of stars and clouds,
expansion, streaming in spiral arms have to be included in
consideration as well. The velocities of stars and gas are
determined (mainly) by the gravitational potential, generated by
the galactic mass distribution. The rotation curve in the galactic
plane is usually determined almost exclusively from radial
velocities (due to very uncertain data on proper motions) of
chosen (distant) tracers with the assumption they do move on
circular orbits around the galactic center. The latter assumption
finds strong, observational support in our case (Fig.
\ref{circle}). Until now, circular motion has been assumed without
evidence because no other assumption was sufficient for modeling
galactic rotation. Our observational evidence is an independent
test showing that the assumption seems well grounded in empirical
data for our choice of tracer.

We adopt the assumption of circular orbits in the galactic
gaseous disc (see Fig. \ref{scheme}); the radial velocities
$V_{\mathrm{rad}}$ towards the observer of
objects situated near the Galactic plane can be
transformed into the circular rotation velocity by inverting
  the following relation (derivable with the sine rule),
\begin{equation}
    V_{\mathrm{rad}} = (\omega_\sun - \omega) R_\sun \sin l \cos b,
    \label{e-vrot-CaII}
  \end{equation}
  where $\omega$$_{\sun}$  is the angular velocity at the Sun
position, i.e. ($V_{\sun}/R_{\sun}$), $\omega$ is the
angular velocity at the position of the considered object
($V_{\mathrm{rot}/R_{*}}$), $R_{\sun}$ is the galactocentric
distance of the Sun, $V_{\sun}$ and $V_{\mathrm{rot}}$ are the rotation
linear velocities of the Sun and the considered star while $l$ and $b$
are the galactic longitude and latitude of the object
respectively. The galactocentric distance of the object $R_{*}$ is
determined from the cosine formula
\begin{equation}
    R_* = \sqrt{R_\sun^2 + (d \cos b)^2 - 2 R_\sun d \cos b \cos l}
    \label{e-Rstar-CaII}
  \end{equation}
  where $d$ is the distance of the object from the Sun.
The distance $d$ can be
estimated from the column density of Ca{\sc ii} for the translucent
clouds (from H and K lines seen in spectra of young stars). Both
the above formulae allow the construction the rotation curve in two
forms: $\omega$ vs. $R_*$ and $V_{\mathrm{rot}}$ vs. $R_*$. In our
approach, instead of the $\omega$ vs $R_*$ or $V_{\mathrm{rot}}$ vs.
$R_*$, we will check the shape of the rotation curve directly on
the observed $V_{\mathrm{rad}}$ vs. $d$ plane. This allows to use
of almost ``raw'' measurements; to
obtain the parameters shown in Fig.~1, one
must convert the radial velocities to orbital ones using
the above formulae and assuming certain values for ${R_\sun}$
and ${V_\sun}$---both of them known to an accuracy of no better than
25\%.

The most promising approach after assuming circular rotation
seems to be to test the rotation curve in the sector of the Galactic
disk around $l=135${\degr}, available to observations from the Northern
hemisphere. We have collected a big sample of high
resolution spectra inside this sector (50 OB stars), some of them
very distant. A very simple model, assuming ${R_\sun}$ and
${V_\sun}$ as stated above (Sofue et al.,  2009), allows evaluation of the
relation between the distances along this sightline and the
expected radial velocities. We need to penetrate
the galactic disk sufficiently deeply with our ``milestones''.

Assuming ${R_\sun}$ = 8.0 kpc and
${V_\sun}$ = 210 km/s,
we used the above equations to draw the relations
of $V_r$ vs. $d$ for the
``flat'' and ``Keplerian'' rotational curves for this selected
galactic longitude.

\section{Observations}

As stated above, we selected the directions with galactic
longitudes 135{\degr} and 180{\degr} for observations; the objects
used for our test are close to the galactic equator, being
situated not farther than 100pc from the galactic plane.

Spectra of the selected objects have been obtained during several
runs spanning the period 1999--2014, using  the Bohyunsan BOES (b) and
Terskol MAESTRO (t) echelle spectrographs. The BOES echelle
spectrograph (Kim et al., 2007) is attached to the 1.8m telescope
of the Bohyunsan Observatory in Korea. The spectrograph has three
observational modes providing resolving powers of 30,000, 45,000
and 90,000. The lowest resolution, enabling observation of quite faint,
heavily reddened, distant objects was used in most cases. In
any mode, the spectrograph allows the whole spectral range
from $\sim$3500 to $\sim$10,000 \AA, divided into 75--76 spectral
orders, to be recorded. The selected stars are listed in Tables
\ref{starsN} and \ref{starsant} together with their galactic
coordinates, distances, and orbital and heliocentric radial
velocities.

MAESTRO (MAtrix Echelle SpecTROgraph) is a three branch
cross-dispersed echelle spectrograph installed at the coude focus
(F/36) of the telescope ZEISS-2000 at the Terskol Observatory
(Caucasus mountains). It was designed for stellar spectroscopy
with resolutions from 45,000 to 190,000 in the spectral range
3,500--10,000 \AA. Using the lowest resolution mode (sufficient
for our programme) one can reach spectra of objects as faint as
$\sim$10$^m$ with a signal-to-noise (S/N) ratio not less than 70, sufficient for measurements of Ca{\sc ii} line.

Two spectra, depicted in Fig.~\ref{idea} are high resolution
spectra, acquired using the HARPS spectrograph at the 3.6 m ESO
telescope (HD147889) and HARPS--N at the 3.5 m Telescopio
Nazionale Galileo. These were used here only in the justification of our method, and
not for the measurements presented in this work.

All the spectra were reduced and measured in a standard  way using the packages
IRAF (Tody 1993) and DECH\footnote{\url{http://gazinur.com/DECH-software.html}}.

\section{Results}

We have measured, for all available targets, both distances and
radial velocities using the Ca{\sc ii} H and K lines
(Tables~\ref{starsN} and \ref{starsant}). In every case we used
the radial velocity of the most blue-shifted Doppler component
(Fig. \ref{lines_agr}) to construct the galactic rotation curve.
The precision of the radial velocity is in every case no worse
than 1.3~km/s (for the spectra with $R$=30,000). The accuracy of
distances is likely to be worse; however, because these are based
on a single calibration, they are much more reliable than those
obtained using spectroscopic parallax. The latter suffers because
of systematic uncertainties related to determination of spectral
type (as shown above---Fig.~\ref{klasyk}), variability of
brightness (usually unknown in the case of distant, rarely
observed objects) and the reddening index ($R_V$) of the object
studied.

To compare the rotational curve determined in the way described
above for the thin Galactic Disk with the kinematical properties
of other possible tracers of this structure, we restricted the
choice of ``other'' tracers to open clusters,
which are massive and thus
less susceptible to random gravitational disturbances. Moreover,
open clusters are too young to be incomers from recent mergers.
Our present analysis is thus restricted to the objects situated in
very close to the galactic equator and formed/born in
the places they are observed in.

For the same reason we do not intend to interpret the observed
(published) rotational curves of other galaxies as these curves
obviously present kinematics composed from a mix of many kinds of
tracers: those representing the thin disk as well as those
significantly gravitationally disturbed, which create substantial
scatter in the average rotational curve of our Galaxy (Fig.~1 of
Sofue et al., 2009). Our observations and analyzes are restricted
to the thin disc of our Galaxy. This is also why we do not
relate our results to galactic rotation curves inferred from
evolved, giant stars like those of Xue et al. (2008) or Bovy et
al. (2012) as these results concern other galactic population with larger velocity dispersion,
while our considerations concern exclusively thin disc, young objects.    

\begin{figure}
\includegraphics[angle=270,width=9cm]{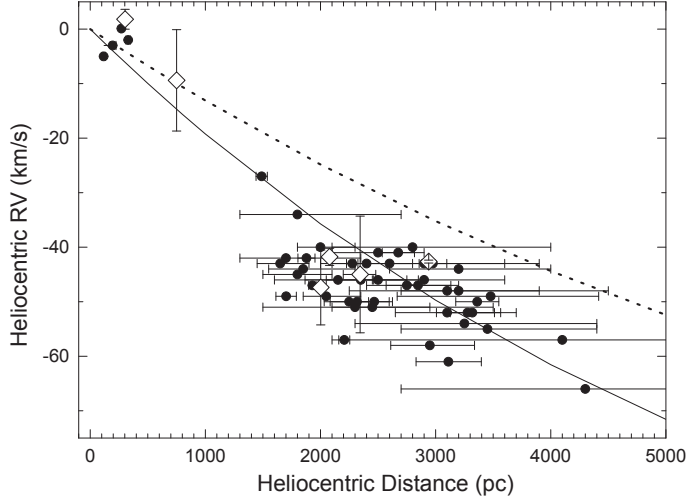}
\caption{Radial velocity curve for the direction $l=135\degr$.
``Flat'' rotation (dotted line) and Keplerian (solid line) theoretical
curves are shown. Filled circles represent our observed objects
(Ca{\sc ii} clouds) and open diamonds represent literature data
for open clusters.}
\label{north-new}
\end{figure}

The results of our measurements are collected in Tables~\ref{starsN} and \ref{starsant}.
Our observed targets are shown as filled circles in Fig.~\ref{north-new}.
In this figure, two model rotation curves for the
Galaxy, i.e. flat and Keplerian, are also shown. The reasonably large
scatter seen in Fig. \ref{north-new} most likely is an
effect of the resolution---not high enough to resolve all
Doppler components. Both the resolution and S/N ratio limit the
accuracy of distance estimates, which are the main source of the rotation
curve uncertainties. In Fig.\ref{north-new} we also show (as
open diamonds) seven open clusters (data from
the literature, see Table~\ref{Cluster}) seen at
$l=135${\degr}. We have selected only the clusters whose radial
velocities were determined using at least 4 objects; for the
vast majority of known clusters, radial velocities are based on 1--3
stars. The positions of distant clusters clearly agree with
those of our Ca{\sc ii} clouds within the observed scatter.

However, the distances to open clusters
are uncertain in many cases. According to the compilation by
Subramanian \& Bhutt (2007) the published distances to e.g. NGC
7245 vary between 1925 and 2800 pc, while those to IC166 cover the
range from 3300 to 4800 pc. These high
uncertainties must be reflected in the rotation curve. The
above mentioned clusters have not been used in our analysis.
Once again, our method of determining distances using Ca{\sc ii}
interstellar lines is more accurate and leads to more homogeneous
results.

\begin{figure}
\includegraphics[angle=0,width=9cm]{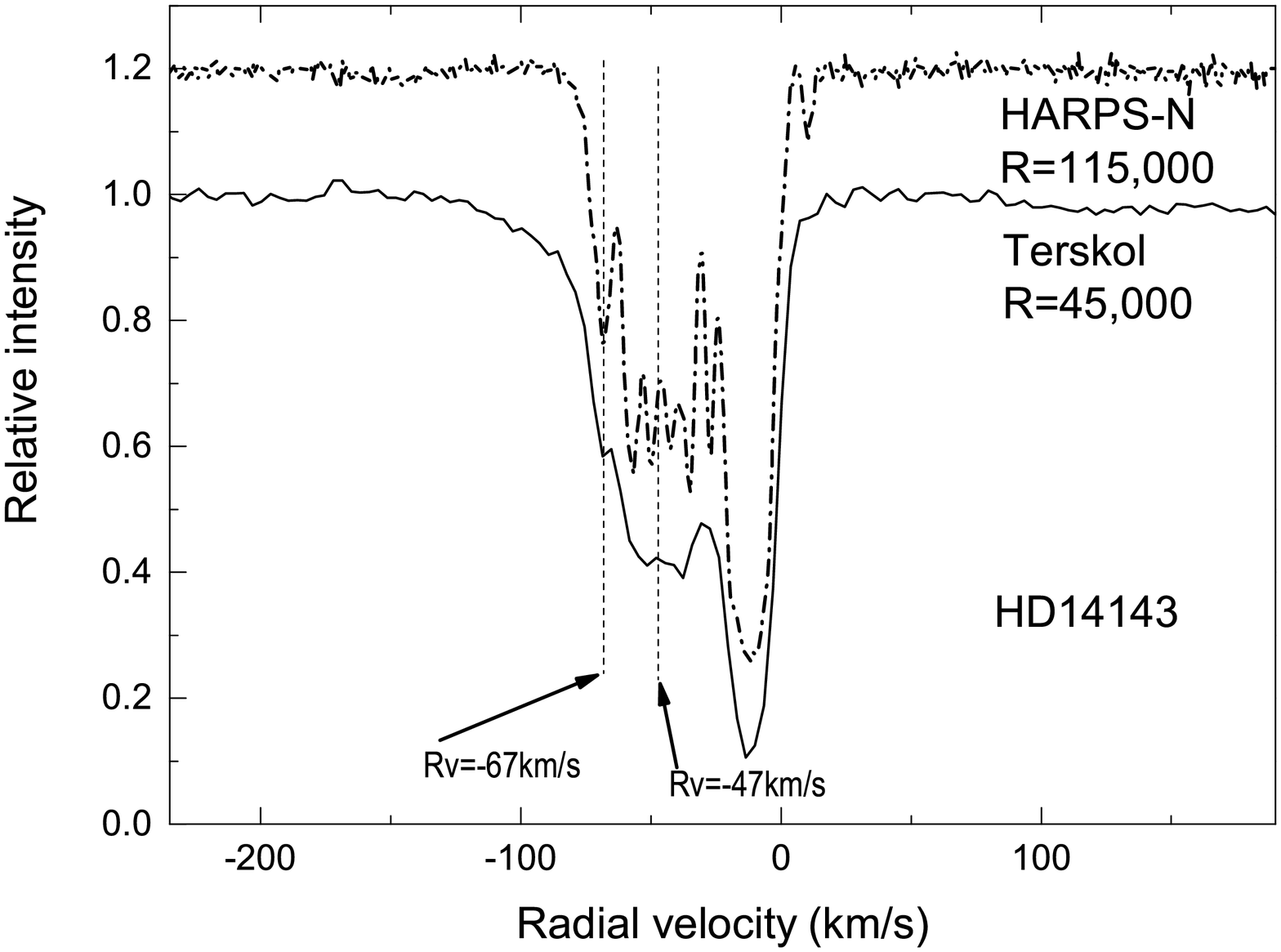}
\caption{Higher resolving power applied to a target in direction $l=135\degr$ reveals more blue component,
i.e. leads to increasing the discrepancy the observed data with the "flat" model (see Fig. \ref{north-new}. }
\label{Reffect}
\end{figure}

Our measurements of radial velocities have, however, uncertainties
related to the relatively low resolution of our spectra. Our first
estimate of the radial velocity of HD2905 (from our spectra) was
$-15$~km/s. We have also checked ultra high resolution results
published by Welty, Hobbs \& Morton (2003). The authors clearly
demonstrate the Doppler component at $-27$~km/s. The latter value
was finally adopted in our work.
 An example that the higher resolving power reveals more blue component in the direction
 $l=135\degr$ is shown in Fig. \ref{Reffect}.
An increase in resolution might influence our results shown in  Fig. \ref{north-new} and Fig.
\ref{com_sofue}. Several very narrow Doppler components might be
found instead of a seemingly single one at lower resolution. This
effect would cause our estimates, the filled circles in Fig.
\ref{north-new} and Fig. \ref{com_sofue}, to be shifted downwards
in the diagrams. Such shifts would make our conclusions even
stronger, as they would move points farther below the flat curve
than they presently lie.

\begin{figure}
\includegraphics[angle=270,width=9cm]{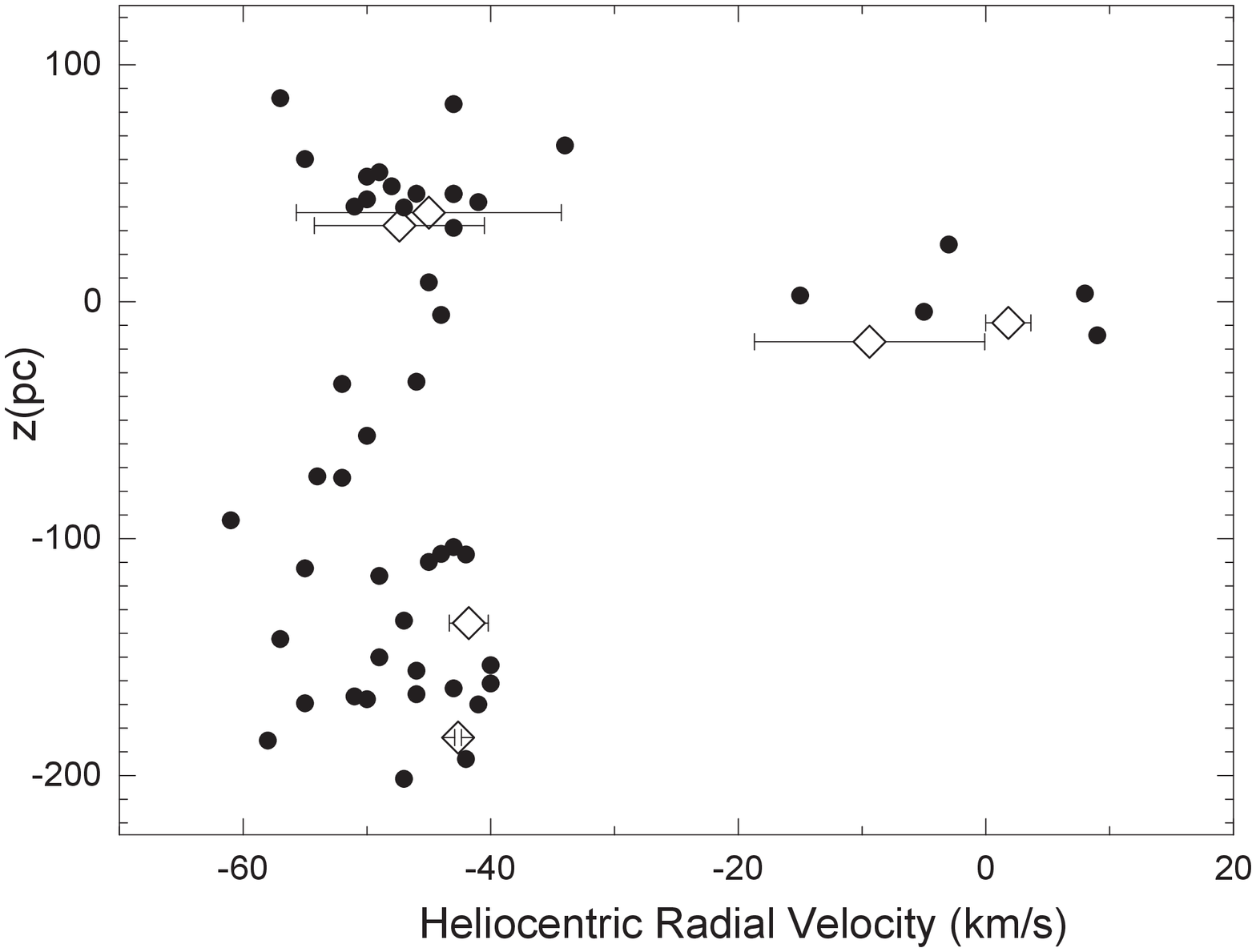}
\caption{Distribution of radial velocities of our targets
in the $l=135\degr$ direction. No systematic effect related to
distance from the galactic plane is observed.}
\label{z}
\end{figure}

\begin{figure}
\includegraphics[angle=270,width=9cm]{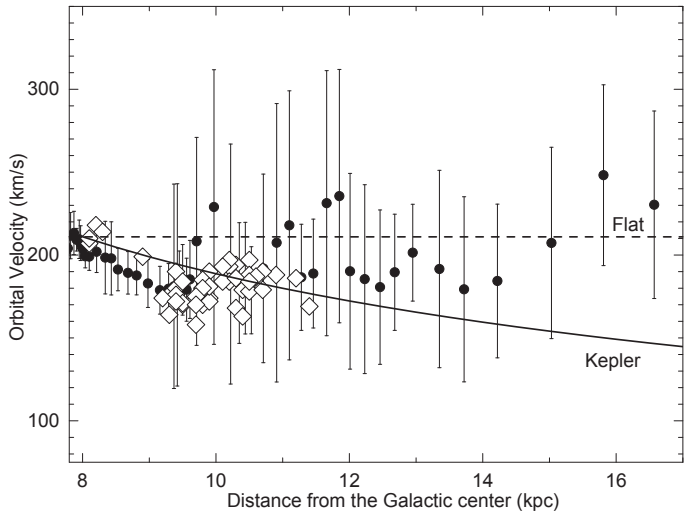}
\caption{Comparison of our measurements (hollow diamonds) with the
rotation curve of our Galaxy compiled by Sofue et al (2009). Our points
fill the very important gap at 9 kpc from the Galactic center. Our
measurements are obviously scattered around the Keplerian rotation
curve. No distant points agree with the flat one.}
\label{com_sofue}
\end{figure}

In the $l=180${\degr} direction (Fig. \ref{circle}), the our data
is scarce but the scatter of radial velocities is very small.
Apparently {\em all} of the observed radial velocities are almost
exactly equal to the solar motion component with respect to the
LSR (almost identical to that of Bovy et al. (2012)), providing
strong evidence that Ca{\sc ii} clouds are orbiting around the
galactic center on circular orbits. In addition, among our
targets, there appears to be no systematic trend of radial
velocities as a function of height from the galactic plane (Fig.
\ref{z}). Apparently our samples represent only the thin galactic
disc.

Equations (\ref{e-vrot-CaII}) and
  (\ref{e-Rstar-CaII}) enable the conversion of our distance
and radial velocity measurements to the plane: distance from the
galactic center versus orbital speed, i.e. we plot the rotation
curve using the same parameters as in Fig.~\ref{sofue}. We
reproduce the set of Sofue et~al. average points and show our
converted measurements. As demonstrated in Fig.~\ref{com_sofue},
our points are scattered around the Keplerian rotation curve as in
Fig.~\ref{north-new}. However, to make the latter plot, we had to
assume certain values for the orbital radius and the speed of the
Sun. These are assumed to be the same as those of Sofue. However,
the orbital radius of the Sun available in the literature ranges
between 7.5 and 10 kpc and the Sun's orbital velocity is reported
between 200 and 250 km/s (Brand \& Blitz (1993)).

We checked the agreement of the postulated rotation curves,
based on different solar orbital velocity and its distance from
the center of the Milky Way with the observed points; the latter
clearly follow the ``Keplerian'' curve. Let's emphasize that the
Sun's specific velocity is almost perpendicular to the l=135
direction and thus its Doppler component along this sightline is
practically zero.

The positions of points in Fig.~\ref{com_sofue} depend on the
assumed values of solar radius and orbital velocity; on the other
hand in Fig.~\ref{north-new} only the theoretical curves depend on
these values, while the points, representing original
measurements, do not. Thus, we consider the plot in
Fig.~\ref{north-new} more reliable. Such a plot may be used to
select the most likely correct values of the orbital parameters of
the Sun (radius of the orbit and velocity).

\section{Discussion} \label{s-discuss}

Let us check whether our result agrees with other measurements,
in particular, with the rotation curve based on spectroscopic parallax distances.
Table \ref{starsSK} contains our targets with the following data: Sp/L (taken from Simbad together with
photometric data), radial velocity of the star, radial velocity of
the CH main features, spectrophotometric distance from the Sun and
the same distance converted to a galacto-centric
distance. As shown in Fig.~\ref{radvels}, the main component of the CH line is usually
closer than the observed star. The galactic rotation curve was in several cases built using CO lines'
velocities and stellar distances. We have collected the published
CO column abundances and related them to our own measurements of
CH abundances (Table~\ref{relcoch}). Both abundances are correlated
in Fig.~\ref{relcoh}. It is evident that the abundances of
the two species are correlated and thus it is reasonable to infer that they are also spatially
correlated and that the radial velocities of the two species should be identical.

However, as shown in Fig. \ref{radvels}, the main component of CH
originates closer to the Sun than
to the observed star. Thus, using spectroscopic distances and CH
radial velocities must lead to an incorrect result. The radial
velocities of the main CH components are much lower than those of
the most blue-shifted ones. This moves points in Fig.
\ref{north-new} up and mimics a flat rotation curve. In fact, the
radial velocities should be measured for the most blue-shifted
components which, in the case of molecular features, are barely
visible in most cases (one can observe them only in very high
S/N spectra).

\begin{figure}
\includegraphics[angle=00,width=9cm]{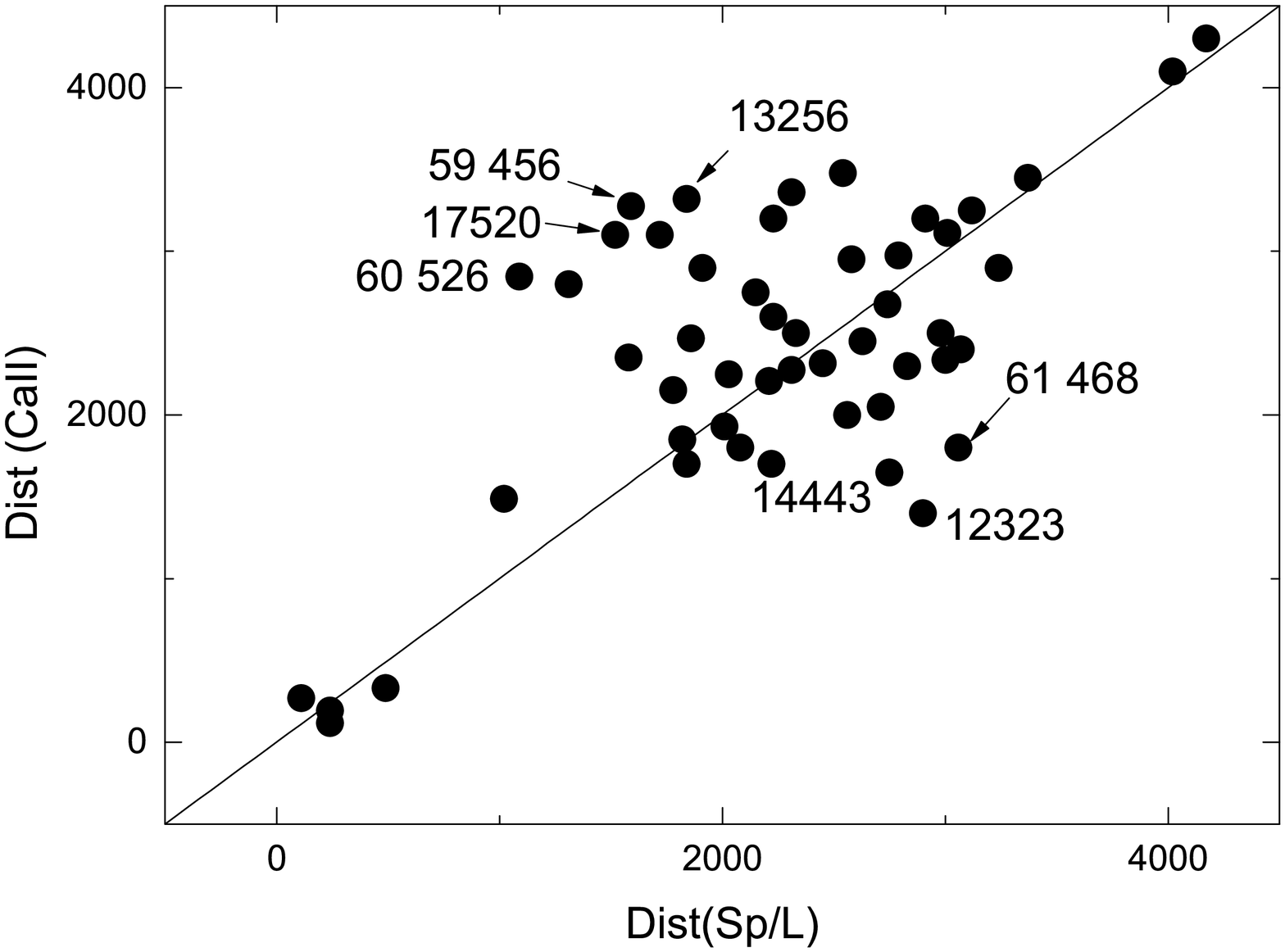}
\caption{Comparison of our Ca{\sc ii} distances with those
from spectroscopic parallax. Some outliers
(see Sect.~\protect\ref{s-discuss})
are marked in the plot.}
\label{odlegl}
\end{figure}

It is also interesting how far our distance estimates coincide
with the spectroscopic ones. Using the data from Tables~\ref{starsN}
and \ref{starsSK}, we have correlated our distances (from
Ca{\sc ii} lines) with the spectroscopic ones (Fig. \ref{odlegl}).
There is a correlation but there are several serious outliers.

\begin{figure}
\includegraphics[angle=00,width=9cm]{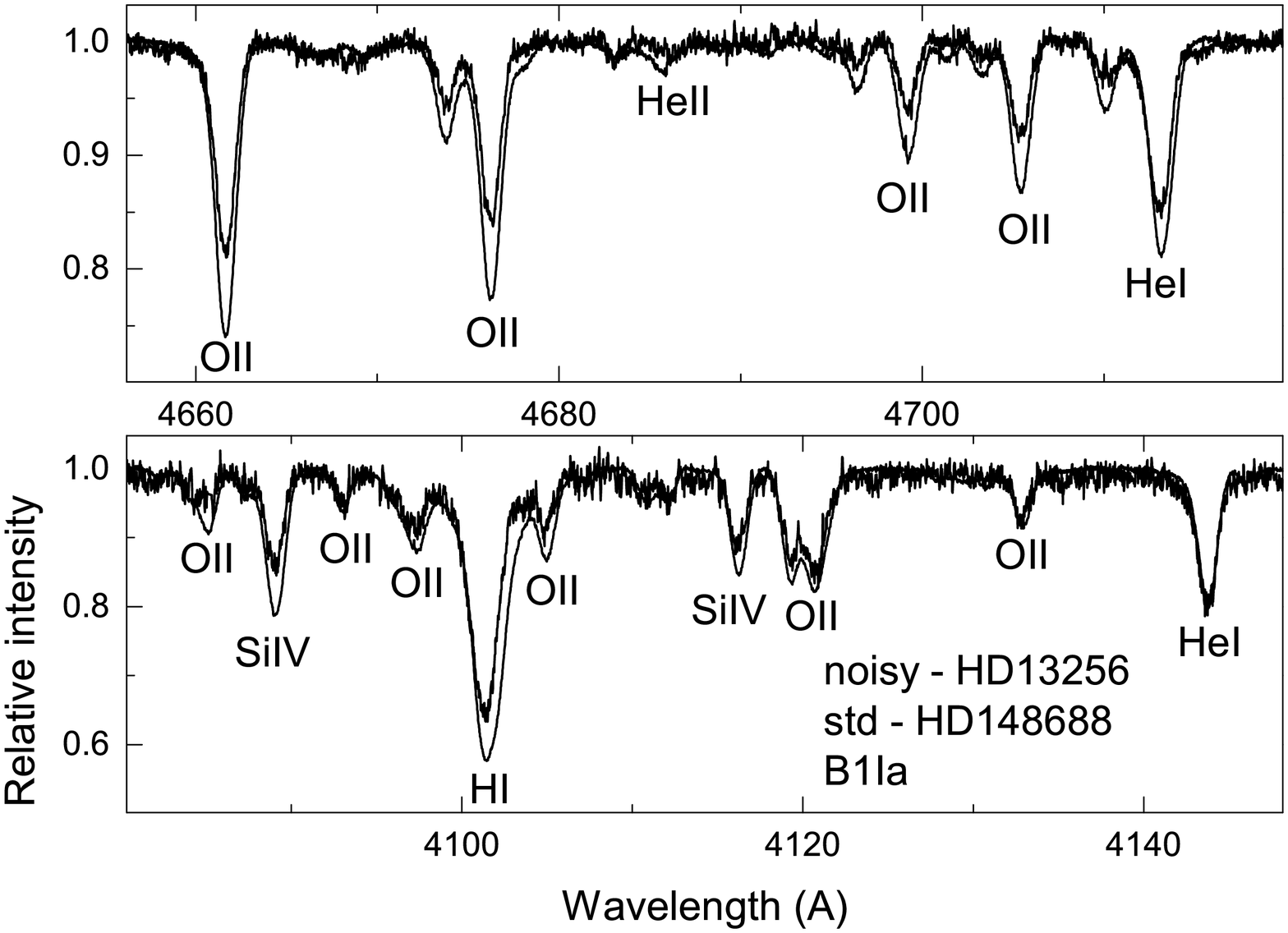}
\caption{Comparison of the spectra of HD13256 (outlier) and
HD148688 (B1Ia standard). HD13256 is likely to be a bit colder
(no HeII line) and of higher luminosity (H{\sc ii} narrower).}
\label{spl13256}
\end{figure}

The published classification of BD+59-456
is evidently flawed. Let us consider another object,
HD13256, for which we have a high quality spectrum. The object is
classified as B1Ia. We compare its spectrum with that of the B1Ia
standard---HD148688 (according to Walborn \& Fitzpatrick (1990)).
Figure~\ref{spl13256} demonstrates that HD13256 is
probably of a later spectral type (since it lacks the HeII line
that is clearly seen in HD148688) and of higher luminosity (the H{\sc i} line
is narrower than in HD148688). Thus, the star may belong to the luminosity class
Ia-0, leading to a much higher spectroscopic distance, which agrees with that
based on Ca{\sc ii} lines.

Generally the highest luminosity OB stars are very scarce and thus
it is very difficult to find two identical spectra which may lead
to identical luminosities. Apparently all the outliers in Fig.
\ref{odlegl} can be ``corrected'' to our Ca{\sc ii} estimates by
means of changing Sp/L's within the commonly accepted errors.
The published Sp/L's are seriously flawed in
many cases and thus it is practically impossible to apply these
values if there is no possibility to check them using
spectra of sufficient quality.

Nevertheless, we
check the rotation curve inferred from
measurements of stellar distances (using
spectroscopic parallax) and of radial velocities based either on
stellar lines or interstellar molecular features---in our case
the CH radical, well correlated with CO, as shown in Fig.
\ref{relcoch}.

\begin{table*}
\caption{Radial velocities based on stellar He{\sc i} lines,
interstellar CH lines and heliocentric distances, based on the
spectroscopic parallax and galactocentric distances. }
\label{starsSK}
\scalebox{0.6}{
\begin{tabular}{rrrrrrr}
\hline
      Star& Sp/L & (B-V) & RV*     & RV(CH) & dist(Sp/L)  & R$_{gc}$ \\
          &      &       & (km/s)  &  km/s  &   (pc)      &   (kpc)  \\
\hline
+56-574t & B1III    & 0.3   & -36  &  -7 &    1840 & 9.39     \\
+59-451b & B1II     & 0.69  & -23  &  -7 &    2450 & 9.84      \\
+59-456b & B0.5V    & 0.55  & -49  &  -6 &    1590 & 9.17     \\
+60-470t & O8V      & 0.70  & -70  & -14 &    2230 & 9.68    \\
+60-493b & B0.5Ia   & 0.79  & -38  &  -8 &    2790 & 10.16     \\
+60-498t & O9.5V    & 0.53  &  -9  & -10 &    2230 & 9.70      \\ 
+60-499t & O9V      & 0.54  & -38  & -11 &    2830 & 10.19      \\ 
+60-501t & O6.5V    & 0.45  & -63  & -15 &    3240 & 10.54     \\
+60-513t & O7.5     & 0.49  & -85  & -14 &    2540 & 9.96    \\
+60-526t & B2III    & 0.64  & -34  &  -8 &    1090 & 8.81     \\
+60-594b & O9V      & 0.36  & -102 & -14 &    2310 & 9.82    \\
+61-411t & O8       & 1.00  & -42  & -21 &    4020 & 11.17    \\
+61-468t & B2III    & 0.23  & -47  &  -7 &    3060 & 10.41     \\
2905t    & B1Ia     & 0.14  &   0  & -20 &    1020 & 8.57      \\
5394t    & B0IV     & -0.15 &   0  &   0 &    240  & 8.13      \\
12323t   & O9V      & -0.02 & -72  & -16 &    3700 & 10.85     \\ 
13256b   & B1Ia     & 1.17  & -47  & -54 &    1840 & 9.35     \\ 
13267t   & B5Ia     &0.33   & ---  & --- &    2580 & 9.95     \\ 
13716t   & B0.5III  & 0.31  & -47  &  -2 &    1820 & 9.35      \\
13758t   & B1V      & 0.32  & -83  &  -3 &    1310 & 8.97     \\ 
13831t   & B0III    & 0.10  & -62  & -32 &    2910 & 10.24     \\ 
13841t   & B2Ib     & 0.25  & -45  & -16 &    2330 & 9.77      \\ 
13854t   & B1Iab    & 0.28  & -32  &  -6 &    1860 & 9.39     \\ 
13866t   & B2Ib     & 0.17  & -51  & -12 &    2710 & 10.08      \\
13890t   & B1III    & 0.19  & -45  & -17 &    2150 & 9.62     \\
13969t   & B1IV     & 0.31  & -65  & -15 &    1580 & 9.17      \\
14014t   & B0.5V    & 0.14  & -45  &  -9 &    1720 & 9.28     \\
14053t   & B1II     & 0.25  & -45  & -19 &    3070 & 10.38     \\
14134t   & B3Ia     & 0.45  & -30  & -13 &    2210 & 9.68      \\ 
14143t   & B2Ia     & 0.50  & -51  & -10 &    2010 & 9.52    \\ 
14302t   & B1II-III & 0.26  & -44  & -44 &    2560 & 9.97      \\ 
14357t   & B1.5II   & 0.32  & -39  & -11 &    2630 & 10.03     \\
14434t   & O6.5     & 0.16  & -90  & -19 &    2980 & 10.33      \\ 
14442t   & O6e      & 0.41  & ---  & --- &    3120 & 10.41     \\  
14443t   & B2Ib     & 0.34  & -37  &  -4 &    2750 & 10.13         \\
14476t   & B0.5III  & 0.38  & -38  &  -4 &    2080 & 9.58          \\
14818t   & B2Ia     & 0.30  & -50  & -13 &    2220 & 9.71          \\
14947t   & O5.5f    & 0.46  & -46  & -10 &    3010 & 10.35         \\
15558t   & O5f      & 0.50  & -85  &  -7 &    2740 & 10.11         \\
15570t   & O4If     & 0.68  & -44  &  -6 &    2310 & 9.77        \\
15629t   & O5Vf     & 0.42  & -40  & -10 &    3370 & 10.64        \\
15785b   & B1Iab    & 0.57  & -32  &  -4 &    3000 & 10.35          \\
16310t   & B1Ib     & 0.65  & -53  &  -8 &    1780 & 9.36          \\
16429t   & O5.5Iab  & 0.62  & -50  &  -5 &    2030 & 9.56         \\
17505t   & O6.5V    & 0.40  & +25  &  -6 &    1910 & 9.49         \\
17520t   & O9V      & 0.32  & -29  &  -7 &    1520 & 9.17        \\
20336t   & B2.5V    & -0.19 & -18  &  -1 &    240  & 8.18        \\
25940t   & B3V      & -0.03 &  +7  &  +7 &    110  & 8.10           \\
236960t  & B1       & 0.46  & -37  &  -8 &    4170 & 11.32         \\
27192t   & B1.5IV   & -0.01 &  +4  &  +4 &    490  & 8.43           \\
\hline
\end{tabular}
}
\end{table*}

\begin{figure}
\includegraphics[angle=0,width=9cm]{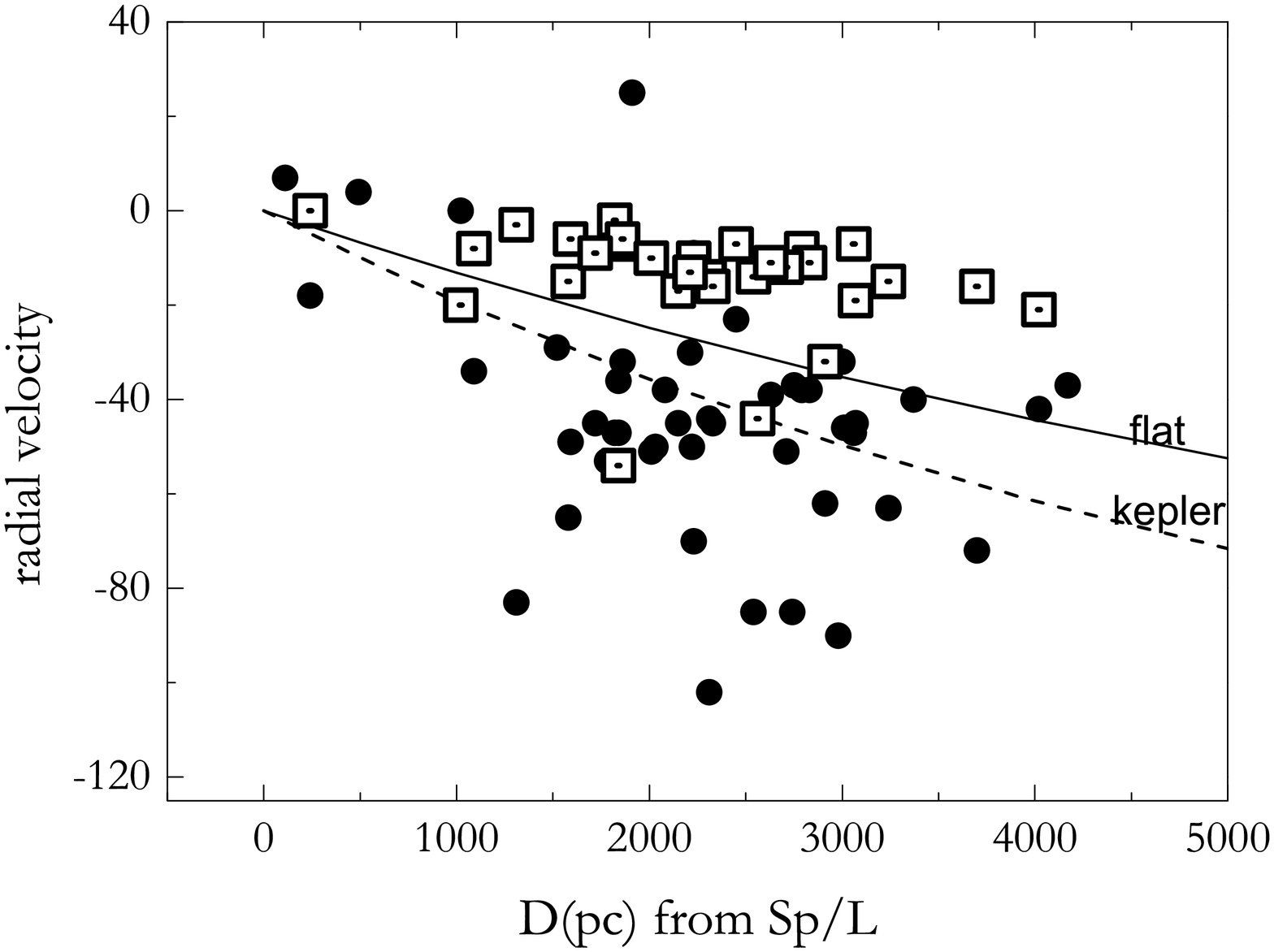}
\caption{Comparison of the rotation curves based on
spectrophotometric distances and radial velocities from
stellar lines (filled circles) or from main CH radial
components (squares)}.
\label{rot_ch_star}
\end{figure}

The result is shown in Fig. \ref{rot_ch_star}. We have plotted two
sets of radial velocities versus spectroscopic parallax distances.
Dots represent radial velocity measurements based on stellar
lines. The result is as expected: the scatter is much higher than
that in Fig. \ref{north-new}. It follows from the phenomena
discussed above: stars are members of multiple systems or of
clusters and thus their radial velocities are sums of the orbital
(around the galactic center) and local components. In spite of
this, the stellar line radial velocities do favor the Keplerian
rotation curve over the flat one, but less clearly than in
Fig.~\ref{north-new}. This demonstrates once again that
measurements of interstellar Ca{\sc ii} lines lead to much more
precise values.

One can express a doubt whether the Ca{\sc ii}
distances are not systematically incorrect due to the declining
gas density outwards the Milky Way disc. However, this should
result in a systematically growing difference between the Ca{\sc ii}
distances and those following the spectroscopic parallax. This is
apparently not the case (Fig. 15). Also, while using stellar
spectroscopic parallaxes we get the same (Keplerian) rotation
curve as while using the Ca{\sc ii} method (Fig. 18, Table 5).

The squares in Fig. \ref{rot_ch_star} represent stellar
(spectroscopic) distances and radial velocities of the main
components of the interstellar CH 4300.3~\AA\ line. As shown in
Fig. \ref{radvels} the main CH components are usually formed in
interstellar clouds situated much closer than the observed stars.
Connecting their radial velocities with stellar distances is
clearly improper. The radial velocities of CH should represent the
most blue-shifted components of the CH lines; however, the latter
are detectable only in very high quality spectra---in the majority
of cases they fall below the detection level. The location of the
squares shows why in some cases the rotation curve is reported to
have rotation velocity growing outwards.

\begin{figure}
\includegraphics[angle=270,width=9cm]{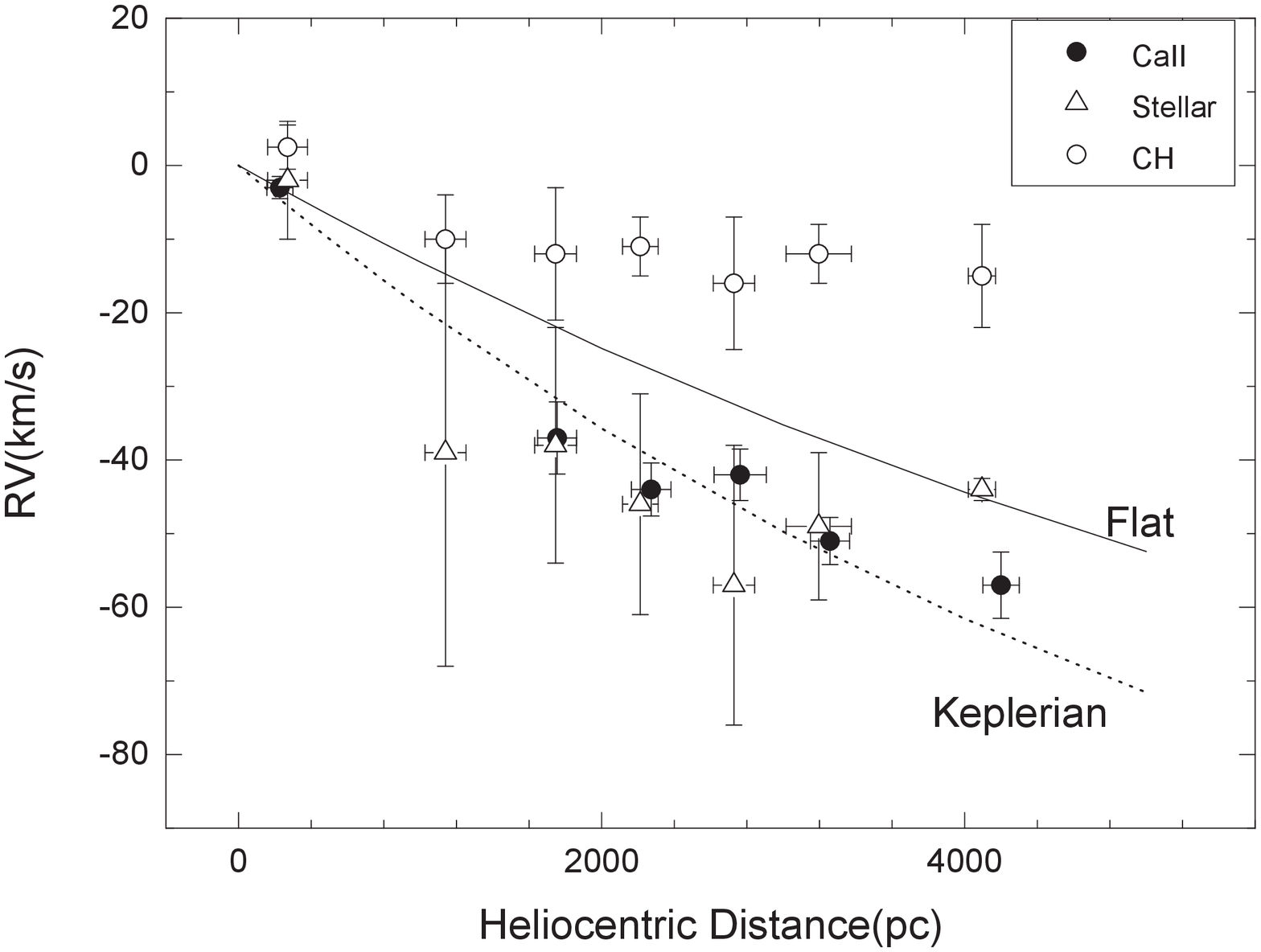}
\caption{Three rotation curves using different methods, averaged
in 500~pc bins. The curve based on Ca{\sc ii} lines is mostly
consistent with that based on stellar distances and radial
velocities. The scatter is much greater in the latter case. The CH
radial velocities may suggest orbital velocities rising outwards,
because molecular clouds are usually much closer than the stars
used to measure distances. The only ``stellar'' point that lies on
the flat curve is based on only two objects. The Ca{\sc ii} point
in the 4000--4500~pc bin is also based on just two objects. }
\label{averages}
\end{figure}

We have averaged our rotation velocities plotted in
Figs~\ref{rot_ch_star} and \ref{north-new} within in 500~pc bins,
and show these in Fig.~\ref{averages}.

Now it is evident that the rotation curve built using traditional
spectroscopic parallax to measure distances to OB stars and
stellar lines to determine radial velocities is consistent with
that based on Ca{\sc ii} lines (apart from the outermost stellar
point, which is based on only two objects). The only obvious
difference is the scatter, being much greater in the case of using
Sp/L distances and stellar line velocities, because of the reasons
explained above. The method based on Ca{\sc ii} lines is the more
precise one.

The results of our distance and radial velocity measurements from
Ca{\sc ii} lines, being more precise than any others and thus best
representing the kinematics of the Galaxy thin gaseous disk,
clearly argue in favor of Keplerian rotation of the latter. We do
not comment on the behavior of other galaxies than the Milky Way
because our method of determining distances using Ca{\sc ii} lines
works properly only inside the thin disk of the Milky
Way---especially along $l = 135${\degr}, because in this
particular direction the range of radial velocities caused by the
galactic rotation is largest, making the Ca{\sc ii} lines
unsaturated even at very large distances. Higher resolution may
lead to discoveries of more Doppler components and thus to larger
shifts of the most blue-shifted Doppler components in the
$l=135${\degr} direction. In that case our points in Fig.
\ref{north-new} could only be moved down, i.e. more strongly
supporting the Keplerian rotation curve.

Moreover, it seems important to extend this method to other
directions in the disk, especially at $l = 225${\degr} (available
from the Southern Hemisphere only) and to extend the galactic
anticentre sample, which presently shows clearly that the radial
velocities of interstellar clouds are all close to zero in
relation to the LSR.

Much bigger samples of targets, observed in the chosen directions,
are needed to make the conclusions well grounded, but the existing
material is sufficient to conclude that the rotation curve of the
thin, gaseous disc of our Galaxy is Keplerian rather than flat.

\begin{acknowledgements}

JK acknowledges the financial support of the Polish National
Science Center during the period 2012--2015 (grant
UMO-2011/01/BST2/05399). GAG acknowledges the support of the
Chilean fund FONDECYT-regular (project 1120190). The authors
acknowledge  Drs. C. Moni Bidin and Boud Roukema for their valuable
comments.

\end{acknowledgements}


\end{document}